\documentclass[11pt,a4paper]{article}

\usepackage{polski}
\usepackage[english]{babel}

\usepackage{jheppub}
\usepackage{amsthm,amssymb,amsmath,epic,float}
\usepackage{rotating,indentfirst,array,varioref}
\usepackage{appendix,tikz,mathtools}
\usepackage{setspace}
\usepackage{longtable}
\usepackage{tikz}
\usetikzlibrary{decorations.pathreplacing}
\usetikzlibrary{calc}
\usepackage{enumitem}
\usepackage{hyphenat}
\usetikzlibrary{positioning}
\usepackage{graphicx}  
\usepackage{adjustbox}

\usepackage{arydshln}
\usepackage{cancel}

\usepackage{comment}
\usetikzlibrary{intersections}
\pgfdeclarelayer{background}
\pgfsetlayers{background,main}

\usepackage{tikz,pgf}
\usetikzlibrary{shapes}
\usetikzlibrary{calc}
\usetikzlibrary{decorations.pathmorphing}
\usetikzlibrary{decorations.pathreplacing,shapes.misc}
\usetikzlibrary{positioning}
\usetikzlibrary{arrows}
\usetikzlibrary{decorations.markings}
\usetikzlibrary{shadings}
\usetikzlibrary{intersections}

\usepackage{tikzsymbols}
\usepackage{pgfplots}
\pgfplotsset{compat=1.18}
\usetikzlibrary{shapes.arrows,arrows.meta,patterns,angles,quotes}
\usepackage{subcaption} 

\usepackage{circledsteps}

\def\be{\begin{equation}}
\def\ee{\end{equation}}

\def\be{\begin{equation}}
\def\en{\end{equation}}
\def\ber{\begin{eqnarray}}
\def\enr{\end{eqnarray}}

\newcommand{\pd}{\partial}
\newcommand{\tr}{\mathop{\mathrm{Tr}}}

\newcommand{\br}[1]{{\overline{#1}}}

\def\<{\left(}
\def\>{\right)}

\usepackage{jheppub}
\makeatletter
\def\@fpheader{\vspace{-.1cm}}
\makeatother

\title{\boldmath WKB-asymptotics for multipoint Virasoro conformal blocks and applications}

\author[a,b]{Aleksandr~Artemev}
\author[a,b,c]{, Dmitry Khromov}

\affiliation[a]{Landau Institute for Theoretical Physics, 142432, Chernogolovka, Russia}
\affiliation[b]{Skolkovo Institute of Science and Technology, 121205, Moscow, Russia}
\affiliation[c]{Moscow Institute of Physics and Technology, 141700 Dolgoprudny, Russia}

\emailAdd{artemev.aa@phystech.edu}

\abstract{We study multipoint Virasoro conformal blocks on the sphere in the comb channel. We arrive at the asymptotic expression for these blocks at large intermediate dimensions, applying WKB method for ``classical BPZ equation'', which is used to study (classical) Virasoro blocks via monodromy method. Several applications of this asymptotic are discussed, such as the possibility to generalize Zamolodchikov's elliptic recursion  and numerical evaluation of amplitudes in minimal string theory. Our expressions pass nontrivial checks, such as agreement with known exact expressions for 5-point blocks in special cases and the usual series expansion of Virasoro blocks computed using AGT correspondence.}


\begin{document} 
\maketitle
\flushbottom
\section{Introduction}
Virasoro conformal blocks, introduced in \cite{bpz}, are important  functions in two-dimensional CFT and string theory. They are the kinematical ``building blocks'' for correlators and are determined completely by Virasoro symmetry. In principle, it means that one can write down an expression for any correlation function in CFT with known spectrum and structure constants. However, due to technical difficulties, apart from some special limits (such as global blocks \cite{Rosenhaus:2018zqn, Ammon:2024axd} or blocks with perturbatively-heavy operators \cite{Alkalaev:2015fbw} --- cases relevant e.\,g. for AdS$_3$/CFT$_2$ correspondence), questions involving  Virasoro blocks on the sphere with more than $4$ external points or higher genus blocks seem to remain relatively unexplored. The present paper intends to establish some results useful for analytic or numerical studies that use multipoint Virasoro blocks in genus zero.

The main motivation for our study concerns theories of ``Liouville gravity''. This is a class of string theories defined by a worldsheet CFT consisting of 3 uncoupled sectors: Liouville theory, BRST-ghosts and some matter CFT, with the total central charge of the 3 sectors equal to zero. Several interesting examples of such theories were conjectured to have a dual description in terms of ``matrix models''/``topological recursion''. This dual approach produces relatively simple and explicit predictions for the values of perturbative string amplitudes. On the level of worldsheet this correspondence is very nontrivial (even when analyzing simple analytic properties, see \cite{Khromov:2025awh}) and explicit checks (numerical or analytic) require a good understanding of the structure of correlation functions in CFT (especially in Liouville theory).

At the moment a lot of numerical studies of the amplitudes involving integration over complex one-dimensional moduli spaces (4-point tachyon amplitude on the sphere and 1-point on the torus) are available in the literature. They include amplitudes in minimal string theory (for $A$-series \cite{Belavin:2010sr, Aleshkin:2016snp} and, more recently, $D$ and $E$ \cite{Rodriguez:2025rte}), Virasoro minimal string \cite{Collier:2023cyw} and complex Liouville string \cite{Collier:2024worldsheet}. The only calculation of higher-dimensional moduli space integrals that we are aware of is the 2-point torus amplitude in $c=1$ string performed in \cite{Balthazar:2017mxh}. 

Two main technical steps that become more difficult for higher-point amplitudes are how to compute efficiently higher-point conformal blocks and how to perform the moduli space integral itself. The most straightforward approach to general conformal blocks would be to use the formulas that follow from AGT correspondence \cite{Alday:2009aq, Alba:2010qc}, giving an explicit combinatorial expression for their coefficients. For spherical blocks this gives a series expansion in the cross-ratios, that has a limited radius of convergence. Moreover, one needs a large number of terms in the expansion to get reliable results after integration and computing the coefficients to high orders using AGT formulas takes a lot of time. For 4-point amplitudes it is known that the problem is greatly simplified if the conformal blocks are expanded in elliptic variable $q$. It is related to cross-ratio $z$ by a well-known ``pillow map'' \cite{Zamolodchikov1987ConformalSI, Maldacena:2015iua} $q = \exp \left(-\pi \frac{K(1-z)}{K(z)} \right)$, which realizes four-punctured sphere moduli space $\mathcal{M}_{0,4}$ as a six-fold covering of one-punctured torus moduli space $\mathcal{M}_{1,1}$ (inverse map is also known as ``modular lambda function''). Such an expansion can be efficiently computed using the celebrated Zamolodchikov recursion \cite{Zamolodchikov1987ConformalSI} (see also \cite{Hadasz:2007nt} for $\mathcal{N}=1$ SUSY case and \cite{Hadasz:2009db} for torus one-point block). This recursion is based on asymptotic form of 4-point conformal blocks when internal dimension $\Delta \equiv \frac{c-1}{24} + P^2 \to \infty$:
\begin{equation}\label{eq:4pointAsymptotics}
\mathfrak{F}^{(c)}(P_1, P_2, P_3, P_4;P|z)
\sim
(16q)^{P^{2}}z^{-\frac{c-1}{24}-P_{1}^2-P_{2}^2}(1-z)^{-\frac{c-1}{24}-P_{2}^2-P_{3}^2}\theta_{3}(q)^{-\frac{c-1}{6}-4\sum_{i}P_{i}^2}
\end{equation}
and known analytic structure in $\Delta$. It turns out that in the 4-point case only a few terms in this expansion apart from the leading asymptotic are actually needed to compute string amplitudes with reasonable precision. The main goal of the present paper is to derive expressions similar to \eqref{eq:4pointAsymptotics} and recursive representation for higher-point blocks in genus zero and test them in practice. We note that our recursion representation is different from the ``$h$-recursion'' previously derived in~\cite{CollierRecursion}, where different limit was taken to compute the asymptotics and the block is computed as a series in usual cross-ratios instead of elliptic variables.

The structure of the paper is as follows. In section~\ref{sec2} we recall the standard method used to compute classical ($c\to \infty$) Virasoro conformal blocks for the case of 5-point block at genus zero. It involves solving an auxiliary ODE and finding the values of parameters for which solutions of the ODE have specified monodromy properties. The asymptotics at large internal dimension in this approach follow from WKB-type expansion for the solution of this equation; with some additional approximations, WKB-series can be expressed in terms of elliptic functions. We then check the obtained expressions against available exact results and asymptotics of AGT formula. In section~\ref{sec3} we give a short discussion on how it can be generalized for higher-point conformal blocks on the sphere. In section \ref{sec3.5} we provide geometric interpretation of our results. Section~\ref{sec4} is devoted to applications of these asymptotics to obtain a Zamolodchikov-like elliptic recursion for multipoint blocks and checks of this recursion. Then, we demonstrate how this formulas can be applied in practice for correlators in Liouville gravity. Before the calculation, we shortly introduce the problem and the example we study --- a 5-point amplitude involving one ``ground ring'' operator. Finally, in section~\ref{sec5} we conclude and discuss directions for future work.

\section{WKB-computation for five-point blocks}\label{sec2}
Let us first introduce our notations. We use ``Liouville parametrization'' of the central charge of the CFT in terms of parameter $b$: 
\[
c \equiv 1 + 6Q^2, \qquad Q \equiv b+b^{-1}. 
\]
The semiclassical limit $c \to \infty$ then corresponds to $b\to 0$. The holomorphic conformal dimensions $\Delta$ of primary operators are parametrized with ``Liouville momenta'' $P$: operator $V_P$ has dimension 
\[
\Delta(P) \equiv \frac{Q^2}{4} + P^2.
\]
In this notation degenerate fields with dimensions from the Kac table have 
\[
P = P_{m,n} = \frac{i (m b + n b^{-1})}{2};
\]
they will also be denoted as $V_{m,n}$. External momenta for $n$-point conformal blocks will be denoted $P_1, \dots, P_n$ and internal $P_{i1},\dots, P_{i\,n-3}$. To shorten the notation we will also sometimes denote by $\{P_k\}$ the collection of external momenta and by $\{P_{i\,k}\}$ the collection of internal momenta.

\subsection{Classical conformal blocks via monodromy method}
Now we recall how classical conformal blocks can be computed. Consider a correlation function in Liouville theory with the insertion of the degenerate field $V_{2,1}$. The corresponding Verma module has a singular vector on level $2$ of the form $(L_{-1}^2 + b^2 L_{-2})V_{2,1}$. Factorization over the corresponding submodule (i.\,e. putting this singular vector to zero) after using conformal Ward identites to express the action of Virasoro modes as differential operators yields the celebrated BPZ equation \cite{bpz}
\begin{equation}
\left[\frac{\pd^2}{\pd z^2} + b^2 \sum \limits_{k=1}^n \left(\frac{\Delta_k}{(z-z_k)^2} + \frac{\pd_k}{z-z_k}\right) \right] \langle V_{2,1}(z) \prod \limits_{k=1}^n V_{P_k}(z_i)  \rangle = 0. \label{bpz}
\end{equation}
Being holomorphic, this equation is also valid for all conformal blocks that enter in the decomposition of this correlator. 

Now consider the semiclassical limit, when $b \to 0$. Path integral representation for Liouville correlators suggests that in this limit correlation functions admit asymptotic expansion in $1/c$, or $b^2$, after taking the logarithm: $\langle \prod V_{P_i}(z_i) \rangle \sim \exp \left(b^{-2} S_{\text{cl}}(bP_i,z_i) + \dots \right)$. The leading term $S_{\text{cl}}$ (``classical Liouville action'') is determined by the saddle point in the path integral; it is only affected by operator insertions if their conformal dimensions also scale in the classical limit: $\Delta = b^{-2}\delta$, $b \to 0$ with $\delta$ finite (in this limit momenta $P$ also tend to infinity with $bP$ finite). The same is valid for conformal blocks (this statement was proven in \cite{Besken:2019jyw} for four-point blocks and more recently in \cite{Gerbershagen:2026cam} in more generality): 
\[
\mathfrak{F}^{(b)}(\{P_k\};\{P_{i\,k}\}|\{z_k\}) \sim \exp \left(b^{-2} \mathfrak{F}^{\text{cl}}(\{bP_k\};\{bP_{i\,k}\}|\{z_k\}) + \dots \right).
\]
For $V_{2,1}$ the conformal dimension stays finite in the semiclassical limit. It means that from the path integral perspective the corresponding correlation function is computed by finding the saddle configuration for other $n$ insertions and simply evaluating $V_{2,1}(z)$ on this saddle configuration. All dependence of the correlator on $z$ in this limit is in a pre-exponential factor which is of order $1$ when $c \to \infty$, while the exponent still depends only on $z_k$. For conformal blocks we can write a similar factorized expression
\[
\mathfrak{F}^{(b)}(\{P_k\},P_{2,1};\{P_{i\,k}\}|\{z_k\},z) \sim \Psi(z) \exp \left(b^{-2} \mathfrak{F}^{\text{cl}}(\{bP_k\};\{bP_{i\,k}\}|\{z_k\}) \right).
\]
Substituting this into the BPZ equation, we can ignore the derivatives $\pd_k$ acting on $\Psi$ in comparison with their action on the exponential, which brings down a large factor $b^{-2}$. Thus, such derivatives can be replaced by multiplying by ``accessory parameters'' $c_k =  \frac{\partial\mathfrak{F}^{\text{cl}}}{\partial z_k}$. The BPZ equation becomes a second-order ODE in the variable $z$.   Only $n-3$ unknown ``accessory parameters'' enter this equation; 3 can be expressed in terms of the other using global Ward identities. 

\begin{figure}[h!]\centering
\includegraphics[width=.7\linewidth]{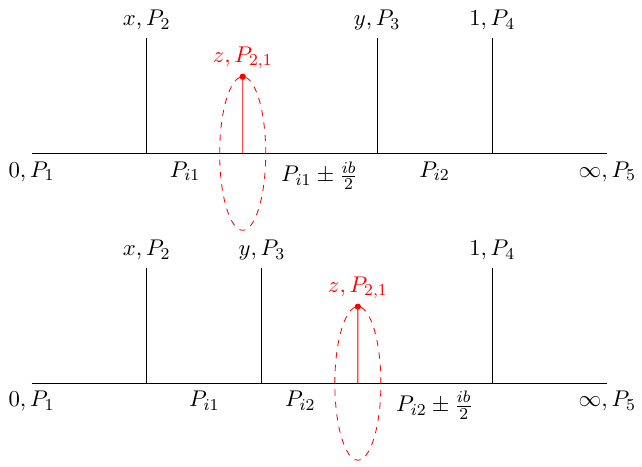}
\caption{Monodromy in the comb channel of the $5$-point block.}
\label{fig:5pointMonodromy1}
\end{figure}

For example, let us specialize to the five-point conformal block on the sphere with insertions $\lbrace P_1,\dots, P_5 \rbrace$ at points $\lbrace 0,x,y,1,\infty \rbrace$. Then the ``classical BPZ equation'' takes the form
\begin{equation}
\Psi''(z) - \left[\frac{x(1-x)c_x}{z(1-z)(x-z)} + \frac{y(1-y)c_y}{z(1-z)(y-z)} -Q(z, x, y)\right]\Psi(z) = 0, \label{5p-cl-bpz}
\end{equation}
where
\begin{equation}
Q(z, x, y) = \frac{\delta_1}{z^2} + \frac{\delta_2}{(z-x)^2} + \frac{\delta_3}{(z-y)^2}+\frac{\delta_4}{(1-z)^2} + \frac{\delta_1 + \delta_2 + \delta_3 + \delta_4 - \delta_5}{z(1-z)}
\end{equation}
includes the terms dependent on external dimensions and, as announced before,
\begin{equation}
c_x(x,y) = \frac{\partial\mathfrak{F}^{\text{cl}}(x,y)}{\partial x}, \qquad c_y(x,y) = \frac{\partial\mathfrak{F}^{\text{cl}}(x,y)}{\partial y}.
\end{equation}
The condition on $c_x$ and $c_y$ such that the solutions of this equation actually correspond to some conformal blocks can be formulated as follows. Solutions $\Psi(z) e^{b^{-2} \mathfrak{F}^{cl}}$ should give 6-point conformal blocks with one degenerate insertion or some linear combination of them. The ``comb channel'' 6-point blocks with degenerate field, illustrated on the diagram~\ref{fig:5pointMonodromy1}, in the classical limit tend to 5-point comb channel blocks and have diagonal monodromy when $z$ goes around $(0,x)$ or $(1, \infty)$, determined by the internal momenta. Instead of finding this diagonal basis for both cases, we can just impose two conditions on the trace of the monodromy matrix for system of solutions, which is basis-independent:
\begin{equation}  
\text{tr\,}M_{(0,x)} = - 2 \cosh 2\pi b P_{i1},\qquad \text{tr\,}M_{(1,\infty)} = - 2 \cosh 2\pi b P_{i2}. \label{monod-con}
\end{equation}
These are two equations that allow to find $c_x$ and $c_y$ as functions of $x$ and $y$. 

\subsection{WKB at large internal momenta: leading orders}
Now we proceed to actually solving these equations. Our goal is to obtain the asymptotic expression for the block when internal dimensions $b^2\Delta(P_{i\,1,2})$  are large compared to external dimensions, which are of order $1$. As we will see later, in this limit accessory parameters $c_x, c_y$ are also of order $(bP_{i\,1,2})^2$, i.\,e. large (this is true in the example of the 4-point conformal block as well). This means that equation \eqref{5p-cl-bpz} is amenable to WKB method: an equation of the form
\begin{equation}
\psi''(z) - \left(\hbar^{-2} t_0 (z) + t_2(z) \right) \psi(z) = 0
\end{equation}
with small parameter $\hbar$ admits a solution  
\begin{equation}
\psi(z) = \exp\left(\int^z p(t, \hbar) dt\right), \quad p(t, \hbar) = \sum_{k = -1}^{\infty} p_k(t) \hbar^k.
\end{equation}
This is not only a formal solution, but in good cases $p(t,\hbar)$ is an asymptotic (Borel summable) series \cite{Iwaki:2025tmq}. Coefficients $p_k$ can be found recursively by substituting the solution into the equation and collecting terms of given order in $\hbar$; in the end they are expressed algebraically in terms of $t_0,t_2$ and their derivatives. The first few are
\begin{equation}
p_{-1}^{\pm}(z) = \pm t_0(z)^{1/2},\,p_{0}^{\pm}(z) = - \frac{1}{4} (\ln t_0(z))',\,p_1^{\pm}(z) = \pm \frac{-\frac{5 t_0'(z){}^2}{16 t_0(z){}^2}+\frac{t_0''(z)}{4 t_0(z)}+t_2(z)}{2 \sqrt{t_0(z)}},\dots
\end{equation}
Choosing a sign for $p_{-1}$ gives two linearly independent solutions $\psi^\pm$. In our case for eq. \eqref{5p-cl-bpz} we can put $x(1-x)c_x$ to be the large parameter $\hbar^{-2}$, so 
$$
t_0(z) = \frac{1}{z(1-z)(x-z)} + \frac{\frac{y(1-y)c_y}{x(1-x)c_x}}{z(1-z)(y-z)},\qquad t_2(z) = -Q(z).
$$
We will then obtain WKB-approximations to $\Psi_{\pm}(z)$ as a series in $1/\sqrt{c_x}$, which, as we expect, later can be transformed into expansion in $1/P_{i\,1,2}$. E.\,g. the first two orders in $\hbar$ give (up to overall normalization, that can be accounted for in the lower integration limit $a$ --- ``normalization point'')
\begin{equation}
\Psi_\pm(z) = \left(\frac{x(1-x)c_x}{z(1-z)(x-z)} + \frac{y(1-y)c_y}{z(1-z)(y-z)}\right)^{-\frac14}\exp \left[\pm\int^{z}_a dt\sqrt{\frac{x(1-x)c_x}{t(1-t)(x-t)} + \frac{y(1-y)c_y}{z(1-z)(y-z)}}\right].
\end{equation}
Solutions normalized at point $a=0$ have diagonal monodromy around $(0,x)$, and for $a=1$ --- around $(1,\infty)$. The conditions \eqref{monod-con} then can be written as
\begin{equation}
\begin{aligned}
\left(\frac{1}{2} \oint \limits_{(0,x)} = \int^{x}_{0}\right) dt \sqrt{\frac{x(1-x)c_x}{t(1-t)(x-t)} + \frac{y(1-y)c_y}{t(1-t)(y-t)}} &= \pi b P_{i1} \equiv \pi b(P+\delta P),\\
\int^{\infty}_{1} dt \sqrt{\frac{x(1-x)c_x}{t(1-t)(x-t)} + \frac{y(1-y)c_y}{t(1-t)(y-t)}}& = \pi b P_{i2} \equiv \pi b P.
\end{aligned}
\end{equation}
Note that the minus sign in \eqref{monod-con} is supplied by the pre-exponential factor\footnote{In this equations, we ignore the possible Stokes phenomena that can happen for general complex values of parameters and affect monodromy matrices. For real $x<y<1$, $P_{i\,k}$ they do not occur; we expect that they should not affect the asymptotic in the general case as well.}. 

The left hand side of these equations is a generic hyperelliptic integral associated to the curve of genus 2 defined by the equation
\begin{equation}
w^2 = z(1-z)(x-z)(y-z)(v-z), \label{hyperel-curve}   
\end{equation}
where
\begin{equation} \label{v-formula-5pt}
v=\frac{x y \left((1-x) c_x+(1-y) c_y\right)}{ x(1-x) c_x+ y(1-y) c_y}  
\end{equation}
is the zero of the integrand. As one can see, $v$ is determined by the ratio of internal momenta $P_{i1}/P_{i2}$. Obviously this can not be solved for $c_x, c_y$ for generic values of large $P_{i\,1,2}$. However, consider the case when $\delta P/P$ is also a small parameter. Two integrals become equal to each other when $c_y \to 0$, so in the limit $\delta P \ll P$ we expect that $c_y \ll c_x$ as well. We can thus expand the LHS in $c_y/c_x$ and then translate this expansion into the one in $\delta P/P$. Note that in the limit $c_y \to 0$ and $v \to y$  the curve degenerates into an elliptic one. All terms in the expansion of the LHS can then be evaluated in terms of elliptic integrals with parameter $x$. The same will be valid in subleading orders in $\hbar$.

One more technical point worth mentioning: when computing the contributions to monodromy from $p_{1}^{\pm}(z)$, we will encounter power-like divergences at the boundaries of integration, since $t_2$ has a double pole at $z=0,x,1,\infty$. They appear because WKB approximation breaks down near these double poles and WKB solution should be ``glued'' with the one accounting for singular terms exactly (for a thorough analysis of such cases in WKB approach see e.\,g. \cite{koikewkb}). In practice this procedure simply means that when computing the monodromy  we shift the boundaries of integration (e.\,g. from $0$ to $x$) by a small $\epsilon$ and discard singular in $\epsilon$ terms.

Now we present the final results of the described procedure --- expansions for accessory parameters in $1/P$ and $\delta P/P$. They can be expressed in terms of standard elliptic integrals $K(x), E(x), \Pi(y,x)$:
\[
c_x = c_{x,-1} b^2P^2 + c_{x,1} + O(P^{-2}), \quad c_y = c_{y,-1}b^2P^2 + c_{y,1} + O(P^{-2})
\]
with
\begin{align}
c_{x, -1} &=  \frac{\pi^2}{4x(1-x)K(x)^2} + \frac{\pi (1-y)(xK(x)+(y-x) \Pi(y,x))}{x(1-x)K(x)^2\sqrt{y(1-y)(y-x)}} \frac{\delta P}{P}
+ \nonumber
\\
&+
\frac{1}{2x(1-x)y(y-x)K(x)^2}
\left[y(y-x)E(x)K(x) + (2x^2(1-y)-y^2(1-x))K(x)^2 + 
\right. \nonumber
\\
&\left.+  4x(y-x)(1-y)K(x)\Pi(y,x) + 2(y-x)^2(1-y)\Pi(y,x)^2\right]\left(\frac{\delta P}{P}\right)^2 + O\left(\frac{\delta P}{P}\right)^3,\\
c_{y,-1} &= -\frac{\pi}{\sqrt{y(1-y)(y-x)}K(x)} \frac{\delta P}{P} - \left(\frac{3x-2y-2xy+y^2}{2y(y-x)(1-y)} + \frac{2\Pi(y,x)}{yK(x)}\right)\left(\frac{\delta P}{P}\right)^2+ O\left(\frac{\delta P}{P}\right)^3,
\end{align}
and
\begin{align}
c_{x,1} &= \frac{1}{4x(1-x)K(x)}\left[3E(x)+(x-2)K(x) - 4E(x)\delta_1-4(E(x)-xK(x))\delta_2 - 4(E(x)-K(x))\delta_4
-\right.\hfill\nonumber
\\
\hfill&-\left.4(E(x)-(1-x)K(x))\delta_5   + \frac{2(y(1-x)K(x)-(y-x)E(x))\delta_3}{y-x}\right] + O\left(\frac{\delta P}{P}\right),\\
c_{y,1} &= \frac{(x-2yx+y(3y-2))}{2y(1-y)(y-x)}\delta_3 + O\left(\frac{\delta P}{P}\right).
\end{align}
What is left is to integrate these formulas with respect to $x$ or $y$ to find the classical conformal block. The result is conveniently expressed in terms of variables
\begin{equation}
q = e^{i\pi \tau(x)}, \quad \tau(x) = i\frac{K(1-x)}{K(x)}, \quad u(y, x) = \frac{\pi}{4K(x)}\int^{\frac{y-x}{x(y-1)}}_0 \frac{dt}{\sqrt{t(1-t)(1-xt)}}.
\end{equation}
and standard Jacobi theta-functions.  $u(y,x)$ can be thought of as a coordinate on the torus with modular parameter $\tau$, or, alternatively, one of the periods of the curve \eqref{hyperel-curve} in the degeneration limit when $v \to y$ (see appendix \ref{sec3-geom-int} for details). 

The terms in the expansion computed above are enough to find the leading terms for the 5-point conformal block at fixed $\delta P$ and external dimensions $\Delta_i$ when $P \to \infty$:
\begin{equation}
\mathfrak{F}^{(b)}(\{P_k\}; P+\delta P, P|x,y) 
=
\mathfrak{f}^{(b)}(\{P_k\}; P+\delta P, P|x,y) \left(1 + O\left(\frac{1}{P}\right)\right)
\end{equation}
with
\begin{multline}\label{eq:asymptotics}
\mathfrak{f}^{(b)}(\{P_k\}; P+\delta P, P|x,y)  = (16q)^{P^2}e^{4iP\delta P (u-i\log 2)} x^{\frac{Q^2}{4}+\delta P^2-\Delta_1-\Delta_2} (1-x)^{\frac{Q^2}{4}+\delta P^2-\Delta_2-\Delta_4}
\times
\\
\times
 (y(1-y)(y-x))^{-\frac12 (\delta P^2+\Delta_3)}
\Theta_3(u|q)^{-4\delta P^2}\theta_3(q)^{3 Q^2 + 6\delta P^2 - 2(2\Delta_1+2\Delta_2+2\Delta_4+2\Delta_5+\Delta_3)} .
\end{multline}
For future reference we note that the term of order $\delta P^2$ can alternatively be written as follows (see useful formulas in appendix \ref{app:formulas}):
\begin{equation}
\left(\frac{x(1-x) \theta_3(q)^6}{\sqrt{y (1-y) (y-x)} \Theta_3(u|q)^4}\right)^{\delta P^2} = \left(\frac{i\Theta'_1(0|q)\theta_2(q) \theta_3(q) \theta_4(q)}{ \Theta_1(u|q)\Theta_2(u|q)\Theta_3(u|q)\Theta_4(u|q)} \right)^{\delta P^2} = \left(\frac{2i \Theta'_1(0|q)}{\Theta_1(2u|q)} \right)^{\delta P^2}.
\end{equation}
A few comments are in order:
\begin{itemize}
    \item the terms independent on $x,y$ can not be found from integrating the accessory parameters. They define the overall normalization of conformal blocks, which will be fixed shortly.
    \item at some terms in the exponents we replaced $b^{-2}$, which would follow from exponentiating the classical conformal block, by $Q^2$. Using our procedure we can not argue rigorously why ``quantum'' corrections of order $1$ and $b^2$ should take this form. However, we believe that such replacement gives an exact asymptotic for quantum block for a few reasons: it makes the expression symmetric under the exchange of $b$ and $b^{-1}$, fixes the coordinate asymptotics at small $x$ and $y$, as well as makes this formula agree with known exact answers in special cases. This will be discussed in the next subsection. 
    \item in this approximation the factor depending on internal dimensions (up to a constant independent on $x,y$) can be written as an exponent of a quadratic form in Liouville momenta
    \begin{equation}\label{eq:5pointPeriods}
    \mathfrak{f}^{(b)} \sim \exp \left(i\pi 
    \begin{pmatrix}
      \delta P  & P\\
    \end{pmatrix}
    \begin{pmatrix}
      \frac{1}{\pi i}\log \frac{\Theta'_1(0|q)}{\Theta_1(2 u|q)}  & \frac{2}{\pi} u\\
      \frac{2}{\pi} u  & \tau \\
    \end{pmatrix}
        \begin{pmatrix}
     \delta P\\
     P \\
    \end{pmatrix}
    \right).
    \end{equation}
    This quadratic form can be compared with the period matrix of WKB curve \eqref{hyperel-curve} in the degeneration limit, which is computed in appendix \ref{sec3-geom-int} (equation \eqref{periodmatrixdiv}). In such a limit, one of the diagonal entries of the period matrix diverges, but the ones that are finite agree with the matrix above. Moreover, in our case WKB curve degenerates in such a way that one can unambiguously extract finite coordinate-dependent part of the divergent matrix element. It is natural to take $v - y$ as the cutoff $\epsilon$ in \eqref{periodmatrixdiv}. Using \eqref{v-formula-5pt}, the divergent term in \eqref{periodmatrixdiv} gives
    \begin{equation}
    \frac{1}{\pi i}\log(\varepsilon u'(y,x))\approx \frac{1}{\pi i} \log \frac{-c_y y(1-y)(y-x) u'(y,x)}{c_x x(1-x)} \approx \frac{1}{\pi i} \log \left(\frac{\delta P}{P} \frac{u'(y,x)}{\frac{\pi}{4\sqrt{y(1-y)(y-x)} K(x)}} \right).
    \end{equation}
    Using \eqref{uderiv}, we see that coordinate dependence cancels in the divergent term. The second term in \eqref{periodmatrixdiv} that does depend on coordinates coincides precisely with the corresponding matrix element in \eqref{eq:5pointPeriods}.
    
    Thus, up to an infinite constant quadratic form in \eqref{eq:5pointPeriods} coincides with the period matrix of WKB curve \eqref{hyperel-curve} in the degeneration limit. We elaborate on this relation in section \ref{sec3.5}.
\end{itemize}

\subsection{Checks}

\subsubsection{Coordinate asymptotics} \label{subsubsec:CoordAsymp}The first thing we check is the correct behaviour in the limit $x\to 0$, $y\to 0$, $y>x$. Using
\begin{equation}
q(x) = \frac{x}{16} + O(x), \qquad u(y, x) = \frac{i}{2}\log\frac{4y}{x} + O\left(x, \frac{x}{y}\right), \qquad x,y \to 0,~y>x,
\end{equation}
we see that the powers of $x$ and $y$ and the normalization is correct:
\begin{equation}
\mathfrak{f}^{(b)}(\{P_k\}; P+\delta P, P|x,y)
=
1\cdot x^{\Delta(P+\delta P) - \Delta_1 - \Delta_2} y^{\Delta(P) - \Delta_3 - \Delta(P+\delta P)}\left(1 + O\left(y, \frac{x}{y}\right)\right).
\end{equation}

\subsubsection{Exact formulas for conformal blocks}
In \cite{FLNO} the authors derived integral representations for discrete families of conformal blocks with one degenerate field $V_{1,2}(y)$. Such conformal blocks satisfy the BPZ equation, which can be transformed into a generalized Lam\'e heat equation
for the function $\Psi(u|q)$ defined by
\begin{multline}
\mathfrak{F}^{(b)}(P_1, P_2, P_{1,2}, P_4, P_5;P\pm i/2b ,P|x,y) 
= 
\\
=
y^{\frac{1}{2b^2}} (1-y)^{\frac{1}{2b^2}} \frac{(y(1-y)(y-x))^{\frac14}x^{-P_1^2}(1-x)^{-P_4^2}}{(x(1-x))^{\frac{2\Delta(P_2)}{3} - \frac13(P_1^2 + P_4^2 + P_5^2)+\frac1{12}}} \frac{(-i\Theta_1(u|q))^{b^{-2}}}{\Theta_1'(0|q)^{\frac{b^{-2}+1}{3}}} \Psi^\pm_P(u|q).
\end{multline}
The prefactor differs from asymptotics $\mathfrak{f}$ by the following factor:
\begin{multline}\label{eq:FLNOpf}
y^{\frac{1}{2b^2}} (1-y)^{\frac{1}{2b^2}} \frac{(y(1-y)(y-x))^{\frac14}x^{-P_1^2}(1-x)^{-P_4^2}}{(x(1-x))^{\frac{2\Delta(P_2)}{3} + \frac1{12}-\frac13(P_1^2 + P_4^2 + P_5^2)}} \frac{(-i\Theta_1(u|q))^{b^{-2}}}{\Theta_1'(0|q)^{\frac{b^{-2}+1}{3}}}
=
\\
=
\frac{\mathfrak{f}^{(b)}(P_1, P_2, P_{1,2}, P_4, P_5; P\pm i/2b, P)}{(16q)^{P^2}e^{\mp2Pb^{-1}(u-i\log 2)}} 
\times
\Theta'_1(0|q)^{\frac{b^2}{3} + \frac{4}{3}(P_1^2+P_2^2+P_4^2+P_5^2)}
\end{multline}
If the momenta $P_k$ take the values
\begin{equation}
P_k = \frac{ib}{4} + \frac{im_k b}{2} + \frac{in_k}{b},\qquad m_k, n_k \in\mathbb{Z}_{\geqslant0}
\end{equation}
integral representations for $\Psi(u|q)$ have been obtained in \cite{FLNO}. In the simplest case $P_1 = P_2 = P_4 = P_5 = ib/4$ the solution with appropriate normalization is simply
\begin{equation}
\Psi_P^{\pm}(u|q) =  (16q)^{P^2}e^{\mp 2b^{-1}P(u-i\log2)}.
\end{equation}
This, together with \eqref{eq:FLNOpf}, means that in this case our asymptotic formula for the conformal block is exact:
\begin{equation}
\mathfrak{F}^{(b)}\left(\frac{ib}{4},\frac{ib}{4}, P_{1,2},\frac{ib}{4},\frac{ib}{4}; \left.P\pm \frac{i}{2b}, P\right|x,y\right) 
=
\mathfrak{f}^{(b)}\left(\frac{ib}{4},\frac{ib}{4}, P_{1,2},\frac{ib}{4},\frac{ib}{4}; \left.P\pm \frac{i}{2b}, P\right|x,y\right). 
\end{equation}
For $P_1 = P_4 = P_5 = ib/4$, $P_2 = ib/4 + imb/2 + in/b$ the solution is (up to some normalization factor)
\begin{multline}\label{eq:PsiSol}
\Psi_P^{\pm}(u|q) \sim \Theta_1'(0|q)^{\frac{2n}{3}\left(1-\frac{2}{b^2}\right)} \int\ldots \int 
\prod_{k=1}^m \left(\frac{\Theta_1(v_k|q)}{\Theta'_1(0|q)^{\frac13}}\right)^{mb^2+2n}\frac{E_1(u+v_k)}{E_1(u)E_1(v_k)}
\times
\\
\times
\prod_{l=1}^n \left(\frac{\Theta_1(v_l'|q)}{\Theta'_1(0|q)^{\frac13}}\right)^{2m+\frac{4n}{b^2}}\left(\frac{E_1(u+v_l')}{E_1(u)E_1(v_l')}\right)^{\frac{2}{b^2}}
\cdot \prod_{i<j} \left|\frac{\Theta_1(v_i-v_j|q)}{\Theta_1'(0|q)^{\frac13}}\right|^{-b^2}  \prod_{i<j} \left|\frac{\Theta_1(v_i'-v_j'|q)}{\Theta_1'(0|q)^{\frac13}}\right|^{-\frac{4}{b^2}}  
\times
\\
\times
\prod_{i,j} \left(\frac{\Theta_1(v_i-v_j'|q)}{\Theta_1'(0|q)^{\frac13}}\right)^{-2}
q^{P^2}e^{\mp 2b^{-1}P(u+b^2(v_1+\ldots +v_m) + 2(v_1' + \ldots + v_n'))} dv_1\ldots dv_m dv_1'\ldots dv_n'.
\end{multline}
Here 
\begin{equation}
E_1(u) = \frac{\Theta_1(u|q)}{\Theta_1'(0|q)}.
\end{equation}
We estimate the behaviour of the integral when $P\to \infty$ using the saddle point approximation. The saddle point equations are
\begin{equation}
\begin{aligned}
& b^2\sum_{i\neq k}\frac{\Theta_1'(v_k-v_i|q)}{\Theta_1(v_k-v_i|q)} + 2\sum_{j=1}^n\frac{\Theta_1'(v_k-v_j'|q)}{\Theta_1(v_k-v_j'|q)} - (mb^2+2n-1)\frac{\Theta_1'(v_k|q)}{\Theta_1(v_k|q)} - \frac{\Theta_1'(u+v_k|q)}{\Theta_1(u+v_k|q)}= \pm 2bP,\\
& \frac{4}{b^2}\sum_{j\neq l}\frac{\Theta_1'(v_l'-v_i'|q)}{\Theta_1(v_l'-v_i'|q)} - 2\sum_{i=1}^m\frac{\Theta_1'(v_i-v_l'|q)}{\Theta_1(v_i-v_l'|q)} - \left(2m+\frac{4n}{b^2}-\frac{2}{b^2}\right)\frac{\Theta_1'(v_l'|q)}{\Theta_1(v_l'|q)} - \frac{2}{b^2}\frac{\Theta_1'(u+v_l'|q)}{\Theta_1(u+v_l'|q)}= \pm 4b^{-1}P.
\end{aligned}
\end{equation}
Since $P\to \infty$ on the right-hand side, the solutions are near zeroes of the denominators of the left-hand side. The saddle point that contributes the most is $v_k \approx 0$, $v_l'\approx 0$, since these are zeroes of $E_1(v_k)$ and $E_1(v_l')$ in the denominator in the integrand. In the leading order in $1/P$ the position of the saddle point is of order $v_k, v_l' \sim 1/P$ and does not depend on $u$ or $q$. Then, since we are not interested in the normalization factor, we just need to evaluate the integrand at the saddle point. We get
\begin{multline}
\Psi_P^{\pm}(u|q) \sim  q^{P^2}e^{\mp 2b^{-1}P u}
\Theta_1'(0|q)^{\frac{2n}{3}\left(1-\frac{2}{b^2}\right)} \prod_{k=1}^m \Theta'_1(0|q)^{\frac23(mb^2+2n)}\prod_{l=1}^n \Theta'_1(0|q)^{\frac23\left(2m+\frac{4n}{b^2}\right)}
\times
\\
\times
\prod_{i<j}^m \Theta_1'(0|q)^{-\frac23 b^2}\prod_{i<j}^n \Theta_1'(0|q)^{-\frac8{3b^2}} \prod_{i=1}^m\prod_{j=1}^n \Theta_1'(0|q)^{-\frac43} 
= 
q^{P^2}e^{\mp 2b^{-1}P u} \Theta_1'(0|q)^{\frac{((2m+1)b + 4nb^{-1})^2}{12} - \frac{b^2}{12}} .
\end{multline}
Comparing this with \eqref{eq:FLNOpf} we see that the function $\mathfrak{f}$ correctly replicates the asymptotics of the exact conformal block.

The solutions for cases when the only momentum not equal to $ib/4$ is not $P_2$ are obtained from \eqref{eq:PsiSol} by replacing all $E_1$ in the integrand in the following way:
\begin{equation}
P_1: E_1 \to E_2 = \frac{\Theta_2(u|q)}{\theta_2(q)}, \quad P_3: E_1 \to E_4 = \frac{\Theta_4(u|q)}{\theta_4(q)}, \quad P_4: E_1 \to E_3 = \frac{\Theta_3(u|q)}{\theta_3(q)}.
\end{equation}
These replacements do not influence the leading order behaviour of the integral, so we still get the same asymptotic, which agrees with \eqref{eq:FLNOpf}.

\subsubsection{AGT formula}
AGT correspondence allows to derive a formula for the $5$-point conformal blocks in the comb channel as a series expansion in $y$ and $x/y$; for reference see \cite{Alba:2010qc}. We want to compare the formula with our $5$-point asymptotics $\mathfrak{f}$. To do that we compare the logarithm of \eqref{eq:asymptotics} with the logarithm of the AGT formula. First few terms of the AGT series are:
\begin{multline}\label{eq:AGT.AGTSeries}
\log \frac{\mathfrak{F}_{\text{AGT}}^{(b)}(\{P_k\}; P+\delta P, P|x,y)}{x^{\Delta(P+\delta P) - \Delta_1 - \Delta_2} y^{\Delta(P) - \Delta_3 - \Delta(P+\delta P)}} = \frac{(Q^2 + 4P^2+4P_4^2-4P_5^2)(Q^2 + 4P_3^2-8P\delta P-4\delta P^2)}{8(Q^2 + 4P^2)}y
+
\\
+
\frac{(Q^2 + 4(P+\delta P)^2 - 4P_1^2 + 4P_2^2)(Q^2 + 4P_3^2 + 8P\delta P+4\delta P^2)}{8(Q^2 + 4(P+\delta P)^2)}\,\frac{x}{y} 
+ 
\\
+
\frac{(Q^2 + 4P^2 +4P_4^2 - 4P_5^2)(Q^2 + 4(P+\delta P)^2 + 4P_2^2 - 4P_1^2)(Q^2 + 4(P+\delta P)^2 + 4P^2 - 4P_3^2)}{16(Q^2 + 4P^2)(Q^2 + 4(P+\delta P)^2)}\,y\cdot \frac{x}{y} 
+
\\
+
 O\left(y^2, \left(\frac{x}{y}\right)^2\right).
\end{multline}
In the first few orders our asymptotic is:
\begin{multline}\label{eq:AGT.AsympSeries}
\log \frac{\mathfrak{f}^{(b)}(\{P_k\}; P+\delta P, P|x,y)}{x^{\Delta(P+\delta P) - \Delta_1 - \Delta_2} y^{\Delta(P) - \Delta_3 - \Delta(P+\delta P)}} = \left(\frac{Q^2}{8} + \frac12 P_3^2 - P\delta P - \frac12 \delta P^2\right)y 
+
\\
+ \left(\frac{Q^2}{8} + \frac12 P_3^2 + P\delta P + \frac12 \delta P^2\right)\frac{x}{y}
+ \left(\frac{Q^2}{16} + \frac14 P_3^2 - \frac{3}{8}P\delta P - \frac18 \delta P^2\right)y^2 
+ 
\\
+
\frac12 \left(\frac{Q^2}{8} + P^2+ (P+\delta P)^2 - P_1^2+P_2^2-\frac12  P_3^2+P_4^2-P_5^2\right)y\cdot\frac{x}{y}
- \frac{P\delta P + \delta P^2}{8}\,y^2 \cdot \frac{y}{x}
+
\\
+
\left(\frac{Q^2}{16} +  \frac14 P_3^2 + \frac{3}{8}P\delta P + \frac{1}{4}\delta P^2\right)\left(\frac{x}{y}\right)^2
+
\frac{P\delta P}{8}\, y
+
\\
+
\frac{1}{64}\left(\frac{11Q^2}{4} + 13P^2+13P\delta P+7\delta P^2 - 14P_1^2+18P_4^2-14P_5^2 + 18P_2^2-7P_3^2\right)y^2\cdot \left(\frac{x}{y}\right)^2 
+ 
\\
+ O\left(y^3, \left(\frac{x}{y}\right)^3\right)
\end{multline}
While the series coefficients are different, if we expand \eqref{eq:AGT.AGTSeries} in $P\to \infty$ and keep the nonvanishing terms, they will coincide with \eqref{eq:AGT.AsympSeries}. Higher order AGT series coefficients quickly become too big to write here and even to compute, but we checked that in the limit $P\to \infty$ the difference of logarithms is indeed zero up to terms including $y^3, (x/y)^3$.


\section{Generalization to spherical $n$-point blocks}\label{sec3}
\begin{figure}[h!]\centering
\includegraphics[width=.8\linewidth]{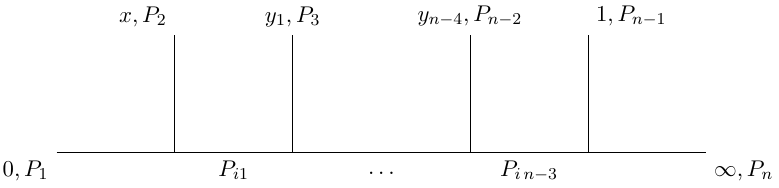}
\caption{Comb channel for the $n$-point conformal block.}
\label{fig:NpointComb}
\end{figure}

Now we consider the same classical BPZ equation we had in section \ref{sec2} but for $n$-point conformal blocks:
\begin{equation}
\Psi''(z) - \left[\frac{x(1-x)c_x}{z(1-z)(x-z)} +\sum_{k=1}^{n-4}\frac{y_k(1-y_k)c_{y_k}}{z(1-z)({y_k}-z)} -Q(z, x, \{y_k\})\right]\Psi(z) = 0, \label{5p-cl-bpz-2}
\end{equation}
where
\begin{equation}
Q(z, x, \{y_k\}) = \frac{\delta_1}{z^2} + \frac{\delta_2}{(z-x)^2} + \sum_{k=1}^{n-4}\frac{\delta_{k+2}}{(z-y_k)^2}+\frac{\delta_{n-1}}{(1-z)^2} + \frac{\sum_{i=1}^{n-1}\delta_i  - \delta_n}{z(1-z)}
\end{equation}
and
\begin{equation}
c_x(x,\{y_k\}) = \frac{\partial\mathfrak{F}^{\text{cl}}(x,\{y_k\})}{\partial x}, \qquad c_{y_k}(x,\{y_k\}) = \frac{\partial\mathfrak{F}^{\text{cl}}(x,\{y_k\})}{\partial y_k}.
\end{equation}
We search for the solutions that satisfy the monodromy conditions
 \begin{equation}
\tr M_{(y_k,\infty)} = -2\cosh2\pi bP_{i\,k}, \qquad 
\tr M_{(1,\infty)} = -2\cosh2\pi b P_{i\,{n-3}}
\end{equation}
for all $k = 1, \ldots, n-4$. Such solutions correspond to conformal blocks in the comb channel (see figure \ref{fig:NpointComb}).

Our goal is to obtain an asymptotic expression for the solutions in the limit when all $bP_{i\, k}$ are simultaneously large compared to external dimensions $\delta_i$. Applying the same procedure as in section \ref{sec2} we get the following monodromy equations:
\begin{equation}\label{eq:monodromyNpoints}
\begin{aligned}
\frac12\int_{\gamma_k} dt \sqrt{\frac{x(1-x)c_x}{t(1-t)(x-t)} + \sum_{k=1}^{n-4}\frac{y_k(1-y_k)c_{y_k}}{t(1-t)(y_k-t)}} &= \pi b (P_{i\,k} -P_{i\,k+1} ) \equiv \pi b \delta P_k,\\
\int^{\infty}_{1} dt \sqrt{\frac{x(1-x)c_x}{t(1-t)(x-t)} + \sum_{k=1}^{n-4}\frac{y_k(1-y_k)c_{y_k}}{t(1-t)(y_k-t)}}& = \pi b P_{i\,n-3} \equiv \pi b P.
\end{aligned}
\end{equation}
In the left-hand side we are integrating a holomorphic differential on a hyperelliptic genus~${g = n-3}$ complex curve
\begin{equation}
w^2 = z(1-z)(x-z)\prod_{k=1}^{n-4}(y_k-z)(v_k-z).
\end{equation}
Here $v_k$ are solutions for $t$ of
\begin{equation}
x(1-x)c_x  + \sum_{k=1}^{n-4}y_k(1-y_k)c_{y_k} \frac{x-t}{y_k-t} = 0, \qquad |v_1|\leq \ldots \leq |v_k|.
\end{equation}
Each contour $\gamma_k$ encircles the interval $(y_k, v_k)$ in the clockwise direction. 

We consider the case when $\delta P_i/P \ll1$ are small parameters of the same order $\delta P/P$. As in the $5$-point case we expect $c_x\sim b^2P^2$ and $c_{y_k}/c_x\sim \delta P_k/P$. Then in the limit $\delta P_k \to 0$ the solution is $c_{y_k} \to 0$ and hence $v_k \to y_k$, so the hyperelliptic curve degenerates. The first correction to $v_k$ is
\begin{equation}
v_k = y_k - \frac{c_{y_k}}{c_x} \frac{y_k(1-y_k)}{x(1-x)}(y_k-x) + O\left(\frac{\delta P}{P}\right)^2.
\end{equation}
Using this we compute the integrals over $\gamma_k$. In the leading order only the terms with $c_x$ and~$c_{y_k}$ contribute. At leading order $c_x$ is the same as before and
\begin{equation}
c_{y_k} = -\frac{\pi b^2P\delta P_k}{\sqrt{y_k(1-y_k)(y_k-x)}K(x)}   + O(\delta P^2).
\end{equation}
Substituting this into the last equation in \eqref{eq:monodromyNpoints} we find first corrections to $c_x$
\begin{equation}
c_x = \frac{\pi^2b^2P^2}{4x(1-x)K(x)^2} + \sum_{k=1}^{n-4}\frac{(1-y_k)(xK(x)+(y_k-x)\Pi(y_k,x))}{x(1-x)\sqrt{y_k(1-y_k)(y_k-x)}K(x)^2} \pi b^2P\delta P_k + O(\delta P^2).
\end{equation}
One could derive higher order corrections in $\delta P_k/P$. 

We now turn to corrections from the external dimensions. Since they are of order $1$, the terms of order $\delta P^2$ that we discarded should be of order less than $1$. Hence, from now we consider $\delta P_k \sim 1/P$. This corresponds to fixed differences of internal dimensions rather than momenta. The correction to $c_{y_k}$ is again obtained from the integral over $\gamma_k$:
\begin{equation}\label{eq:cyNpoints}
c_{y_k} = -\frac{\pi b^2P\delta P_k}{\sqrt{y_k(1-y_k)(y_k-x)}K(x)} + \frac{(x-2y_kx+y_k(3y_k-2))}{2y_k(1-y_k)(y_k-x)}\delta_{k+2} + o(1).
\end{equation}
For $c_x$ we get
\begin{multline} \label{eq:cxNpoints}
c_x = \frac{\pi^2b^2P^2}{4x(1-x)K(x)^2} + \sum_{k=1}^{n-4}\frac{(1-y_k)(xK(x)+(y_k-x)\Pi(y_k,x))}{x(1-x)\sqrt{y_k(1-y_k)(y_k-x)}K(x)^2} \pi b^2P\delta P_k 
-
\\
\hfill-
\frac{E(x)\delta_1+(E(x)-xK(x))\delta_2 + (E(x)-K(x))\delta_{n-1}
+(E(x)-(1-x)K(x))\delta_n}{x(1-x)K(x)}
+ 
\\
+ \frac{3E(x)+(x-2)K(x) }{4x(1-x)K(x)}+
\sum_{k=1}^{n-4} \frac{2(y_k(1-x)K(x)-(y_k-x)E(x))}{4x(1-x)K(x)(y_k-x)}\delta_{k+2} + o(1).
\end{multline}
From \eqref{eq:cxNpoints}, \eqref{eq:cyNpoints} we obtain the asymptotics of the conformal block:
\begin{equation}
\mathfrak{F}^{(b)}(\{P_k\}; P + \sum_{k=1}^{n-4} \delta P_k, \ldots, P+\delta P_{n-4},  P|x,\{y_k\}) 
\sim
\mathfrak{f}^{(b)}(\{P_k\}; P + \sum_{k=1}^{n-4} \delta P_k, \ldots, P+\delta P_{n-4}, P|x,\{y_k\})
\end{equation}
with
\begin{multline}\label{eq:asymptoticsNpoints}
\mathfrak{f}^{(b)}(\{P_k\}; P + \sum_{k=1}^{n-4} \delta P_k, \ldots, P+\delta P_k, P|x,\{y_k\}) = (16q)^{P^2}e^{2i\sum_{k=1}^{n-4}P\delta P_k(u_k-i\log 2)} x^{\frac{Q^2}{4}-\Delta_1-\Delta_2} 
\times
\\
\times
(1-x)^{\frac{Q^2}{4}-\Delta_2-\Delta_{n-1}} \prod_{k=1}^{n-4}(y_k(1-y_k)(y_k-x))^{-\frac12 \Delta_{k+2}}
\theta_3(q)^{3 Q^2 - 2(2\Delta_1+2\Delta_2+2\Delta_{n-1}+2\Delta_n+\sum_{k=1}^{n-4}\Delta_{k+2})} .
\end{multline}
Here we denote $u_k \equiv u(y_k,x)$. As in the $5$-point case (see \eqref{eq:5pointPeriods}), we see that the terms with $P\delta P_k$ and $P^2$ are multiplied by the corresponding entries of the period matrix of the degenerate curve (see \eqref{eq:PeriodMatrix}). We elaborate on this appearance of the period matrix in section \ref{sec3.5}.

In terms of internal dimensions this can be rewritten as 
\begin{equation}
\mathfrak{F}^{(b)}(\{P_k\}; P_{i1}, \ldots, P_{i\,n-3}|x,\{y_k\}) 
\sim
\mathfrak{f}^{(b)}_{\Delta}(\{P_k\}; P_{i1}, \ldots, P_{i\,n-3}|x,\{y_k\})
\end{equation}
for $P_{i1}, \ldots, P_{i\,n-3}\to \infty$ with $P_{i\,k}^2-P_{i\,k+1}^2$ fixed and
\begin{multline}\label{eq:asymptoticsNpoints-2}
\mathfrak{f}^{(b)}_{\Delta}(\{P_k\}; P_{i\,1}, \ldots, P_{i\,n-3}|x,\{y_k\}) = (16q)^{P_{i1}^2}e^{2i\sum_{k=1}^{n-4}(P_{i\,k}^2-P_{i\,k+1}^2) (u_k-\frac{\pi}{2}\tau+i\log 2)} x^{\frac{Q^2}{4}-\Delta_1-\Delta_2} 
\times
\\
\times
(1-x)^{\frac{Q^2}{4}-\Delta_2-\Delta_{n-1}} \prod_{k=1}^{n-4}(y_k(1-y_k)(y_k-x))^{-\frac12 \Delta_{k+2}}
\theta_3(q)^{3 Q^2 - 2(2\Delta_1+2\Delta_2+2\Delta_{n-1}+2\Delta_n+\sum_{k=1}^{n-4}\Delta_{k+2})} .
\end{multline}
Here we added the necessary terms of order $\delta P_k^2$ to obtain the correct limit $x, y_1, \ldots, y_k\to 0$:
\begin{equation}
\mathfrak{f}^{(b)}_{\Delta}(\{P_k\}; P_{i1}, \ldots, P_{i\,n-3}|x,\{y_k\}) =  x^{\Delta(P_{i1}) - \Delta_1-\Delta_2}\prod_{k=1}^{n-4}y_k^{\Delta(P_{i\,k+1})-\Delta(P_{i\,k})-\Delta_{k+2}}\left(1 + O\left(y_{n-4}, \ldots, \frac{x}{y_1}\right)\right)
\end{equation}

\section{Geometric interpretation} \label{sec3.5}
\begin{figure}[h!]\centering
		\includegraphics[width=.9\linewidth]{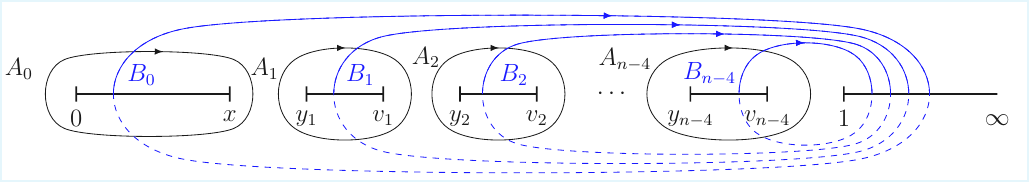}
	\caption{$A$ and $B$ periods on the hyperelliptic curve.}
	\label{fig:NonDegeneratePeriodsN}
\end{figure}

In sections \ref{sec2}, \ref{sec3} we saw that the $P\to\infty$ limit of the conformal block with $\delta P_k/P \ll 1$ fixed is related to the period matrix of the degenerate WKB curve. Here we derive the generalization of this relation for arbitrary $\delta P_k/P$, i.\,e. for the non-degenerate curve. The curve in question is
\begin{equation}
w^2 = z(1-z)(x-z)\prod_{k=1}^{n-4}(y_k-z)(v_k-z) \equiv R(z).
\end{equation}
Here $v_k$ are roots of the polynomial $S(z)$:
\begin{equation}\label{eq:vkDefinition}
S(z) = x(1-x)c_x \prod_{k=1}^{n-4}(z-y_k)  + \sum_{k=1}^{n-4}y_k(1-y_k)c_{y_k} (z-x)\prod_{\substack{l=1\\l\neq k}}^{n-4}(z-y_l) = a^2 \prod_{k=1}^{n-4}(z-v_k).
\end{equation}
We assume that $|x|\leq|y_1|\leq|v_1|\leq\ldots\leq1$ and fix the $A_0, \ldots, A_{n-4}$ and $B_0, \ldots, B_{n-4}$ cycles as shown on the figure~\ref{fig:NonDegeneratePeriodsN}. For convenience we absorb the $\pi^2b^2P^2$ factor into $c_x$, $c_{y_k}$ and temporarily redefine $\delta P_k/P \to \delta P_k$ so that the curve and $c_x, c_{y_k}$ depend only on $\delta P_k$. Then the monodromy equations are
\begin{equation}\label{eq:MonodromyEq-s}
\frac{a}{2} \int_{A_0} \frac{dz}{w} \prod_{i=1}^{n-4} (z-v_i) = 1 + \sum_{k=1}^{n-4}\delta P_k, \qquad \frac{a}2 \int_{A_k} \frac{dz}{w} \prod_{i=1}^{n-4} (z-v_i) = \delta P_k,
\end{equation}
where $k = 1, \ldots n-4$. 
\\
Let $\boldsymbol{\eta}$ be the $g = n-3$ dimensional vector of holomorphic differentials:
\begin{equation}
\boldsymbol{\eta} = 
\begin{pmatrix}
\eta_0
\\
\ldots
\\
\eta_{n-4}
\end{pmatrix} 
= 
\frac{\boldsymbol{\mathcal{U}}(z)}{w}dz, \quad 
\mathcal{U}_0 = \prod_{k=1}^{n-4} (z-v_k), \quad \mathcal{U}_k = \prod_{\substack{l=1\\l\neq k}}^{n-4} (z-v_l).
\end{equation}
Denote by $\boldsymbol{\omega}$ and $\boldsymbol{\omega'}$ the $A$-period and $B$-period matrices of the $\boldsymbol{\eta}$ differentials:
\begin{equation}
\omega_{jk} = \frac12\int_{A_k} \eta_j, \qquad \omega'_{jk} = \frac12\int_{B_k} \eta_j.
\end{equation} 
Then the period matrix of the curve is $\boldsymbol\tau = \boldsymbol{\omega}^{-1}\boldsymbol{\omega'}$. The key ingredient we need for the derivation is the following formula for the derivative of the period matrix (eq-n (4.17) in \cite{EnolskiPeriods}, also see \cite{Thomae1869}): 
\begin{equation}\label{eq:DerivativeOfTau}
\frac{\partial \boldsymbol{\tau}}{\partial e_l} = \frac{i\pi }{R'(e_l)} \boldsymbol{\omega}^{-1} \boldsymbol{\mathcal{U}}(e_l)\boldsymbol{\mathcal{U}}(e_l)^T (\boldsymbol{\omega}^T)^{-1}.
\end{equation}
Here $e_l$ are roots of the polynomial $R(z)$ that defines the curve. A crucial feature of this formula is that the right hand side does not contain explicitly the $B$-period matrix $\boldsymbol{\omega'}$. We also note that \eqref{eq:DerivativeOfTau} is invariant under the change of basis of the holomorphic differentials. Our goal is to prove the following:
\begin{equation}\label{eq:ToProve}
\boldsymbol{\delta P}^T \frac{\partial\boldsymbol{\tau}}{\partial x} \boldsymbol{\delta P} = -i\pi c_x, 
\quad 
\boldsymbol{\delta P}^T \frac{\partial\boldsymbol{\tau}}{\partial y_k}\boldsymbol{\delta P} = -i\pi c_{y_k}, \quad \boldsymbol{\delta P}^T \frac{\partial\boldsymbol{\tau}}{\partial v_k} \boldsymbol{\delta P} = 0, 
\end{equation}
where $\boldsymbol{\delta P}^T = \begin{pmatrix}
 	1 + \sum_{k=1}^{n-4} \delta P_k &
 	\delta P_1 &
 	\ldots &
 	\delta P_{n-4}
\end{pmatrix}$. To do this we need to compute $\boldsymbol{x}^T = \boldsymbol{\delta P}^T \boldsymbol{\omega}^{-1}$. This vector is the solution of the system of equations $\boldsymbol{\delta P}^T = \boldsymbol{x}^T \boldsymbol{\omega}$. From \eqref{eq:MonodromyEq-s} we know that the first row of the period matrix is $a^{-1} \boldsymbol{\delta P}^T$. Hence, the solution is just $\boldsymbol{x}^T = \begin{pmatrix}
a & 0 & \ldots & 0
\end{pmatrix}$. Substituting this into \eqref{eq:DerivativeOfTau} we get
\begin{equation}
\boldsymbol{\delta P}^T \frac{\partial\boldsymbol{\tau}}{\partial e_l} \boldsymbol{\delta P}
=
\frac{i\pi}{R'(e_l)} \boldsymbol{x}^T\boldsymbol{\mathcal{U}}(e_l) \boldsymbol{\mathcal{U}}(e_l)^T \boldsymbol{x} 
= 
\frac{i\pi}{R'(e_l)} a^2\mathcal{U}_0(e_l)^2 = \frac{i\pi}{a^2R'(e_l)} S(e_l)^2.
\end{equation}
From \eqref{eq:vkDefinition} we see that
\begin{equation}
S(x) = x(1-x)c_x\prod_{l=1}^{n-4} (x-y_l),  \quad S(y_k) = y_k(1-y_k)c_{y_k}(y_k-x) \prod_{\substack{l=1\\l\neq k}}^{n-4}(y_k-y_l),\quad S(v_k) = 0.
\end{equation}
Finally, 
\begin{equation}
\begin{aligned}
& a^2R'(x) = \underbrace{a^2 \prod_{l=1}^{n-4}(x-v_l)}_{S(x)} \times \underbrace{x(x-1) \prod_{l=1}^{n-4}(x-y_l)}_{-S(x)/c_k} = - \frac{S(x)^2}{c_x},\\
& a^2R'(y_k) = a^2(y_k-v_k) \prod_{\substack{l=1\\l\neq k}}^{n-4}(y_k-v_l) \times y_k(y_k-1)(y_k-x) \prod_{\substack{l=1\\l\neq k}}^{n-4}(y_k-y_l) = - \frac{S(y_k)^2}{c_{y_k}}.
\end{aligned}
\end{equation}
This gives us \eqref{eq:ToProve}. Returning to our original notation we get
\begin{equation}
i\pi b^2 P^2\boldsymbol{\delta P}^T \frac{\partial\boldsymbol{\tau}}{\partial x} \boldsymbol{\delta P} = c_x, 
\quad 
i\pi b^2 P^2 \boldsymbol{\delta P}^T \frac{\partial\boldsymbol{\tau}}{\partial y_k}\boldsymbol{\delta P} = c_{y_k}, \quad \boldsymbol{\delta P}^T \frac{\partial\boldsymbol{\tau}}{\partial v_k} \boldsymbol{\delta P} = 0.
\end{equation} 
From these equation we get the following result for the asymptotics of the conformal block with zero external dimensions:
\begin{equation}
\mathfrak{F}^{(b)}\left(\left\{\frac{iQ}{2}\right\};\left. P + \sum_{k=1}^{n-4} \delta P_k, \ldots, P+\delta P_{n-4}, P \right|x, \{y_k\}\right) = e^{ i\pi P^2\boldsymbol{\delta P}^T \boldsymbol{\tau}\boldsymbol{\delta P}}\left(1 + O\left(\frac{1}{P}\right)\right).
\end{equation}
This formula for $\mathfrak{F}$ is very beautiful, but not useful in practice: the period matrix is a complicated function of $v_k$, which in turn are determined by $\delta P_k/P$ in a complicated way. It is only analytically tractable in degeneration limits which we consider in other parts of the work.

This result is remarkably similar in spirit to the result of Zamolodchikov on conformal blocks in Ashkin-Teller model \cite{ZAMOLODCHIKOV1987481}. However, in Zamolodchikov's result the roots of the polynomial that defines the curve include only the points of operator insertions, in contrast to WKB curve that has additional branching points $v_k$.

\section{Applications}\label{sec4}

\subsection{$\Delta$-recursion for five-point blocks} \label{ell-recursion}

\subsubsection{Derivation of the recursion}
Perhaps the main application of the large $P$ asymptotics of the $4$-point conformal block is the derivation of Zamolodchikov's recursion relations. In order to do the same thing for the $5$-point block, we should divide it by large $P$-asymptotics to obtain a meromorphic function without essential singularity at infinity. If we tried to use \eqref{eq:asymptotics} we would run into problems: the function $\mathfrak{f}$ in \eqref{eq:asymptotics} involves terms with $\delta P^2$, which are not expressed uniquely and analytically in terms of the internal dimensions. Moreover, the conformal blocks in the residues also would not be analytic functions of $\Delta(P+\delta P)$ and $\Delta(P)$. In order to avoid these problems we consider the conformal block with fixed difference of internal dimensions $\delta\Delta = P_{i1}^2 - P_{i2}^2$, instead of $\delta P$. This means that
\begin{equation}
\delta P \sim \frac{1}{P}
\end{equation}
and we should drop the unwanted terms with $\delta P^2$. If we define
\begin{equation}
\mathfrak{F}^{(b)}(\{P_k\};P_{i1}, P_{i2}|x,y) 
= 
\mathfrak{f}_{\Delta}^{(b)}(\{P_k\};P_{i1}, P_{i2}|x,y)  \mathfrak{H}^{(b)}(\{P_k\}; P_{i1}, P_{2i}|u,q), \label{elliptic-5ptblock-def}
\end{equation}
where
\begin{multline}
\mathfrak{f}_{\Delta}^{(b)}(\{P_k\};P_{i1}, P_{i2}|x,y)
= 
(16q)^{P_{i1}^2}e^{2i(P_{i1}^2-P_{i2}^2) \left(u-\frac{\pi}{2}\tau+i\log2\right)} x^{\frac{Q^2}{4}-\Delta_1-\Delta_2}(1-x)^{\frac{Q^2}{4}-\Delta_2-\Delta_4} 
\times
\\
\times
 (y(1-y)(y-x))^{-\frac12 \Delta_3}
\theta_3(q)^{3 Q^2 - 2(2\Delta_1+2\Delta_2+2\Delta_3+2\Delta_4+\Delta_3)},
\end{multline}
we have
\begin{equation}
\lim_{\substack{P_{i1}, P_{i2}\to \infty\\ P_{i1}^2 - P_{i2}^2~\text{fixed}}} \mathfrak{H}^{(b)}(\{P_k\};P_{i1}, P_{i2}|x,y)
=
\lim_{\substack{P_{i1}, P_{i2}\to \infty\\ P_{i1}^2 - P_{i2}^2~\text{fixed}}} \frac{\mathfrak{F}^{(b)}(\{P_k\};P_{i1}, P_{i2}|x,y) }{\mathfrak{f}_{\Delta}^{(b)}(\{P_k\};P_{i1}, P_{i2}|x,y)} 
= 1.
\end{equation}
Now we consider the $\mathfrak{H}$ block with internal dimensions $\Delta_{i 1} = \Delta + \delta \Delta$ and $\Delta_{i 2} = \Delta$ as a meromorphic function of $\Delta$. It has two series of simple poles at $\Delta = \Delta(P_{m,n})$ and ${\Delta = \Delta(P_{m,n}) - \delta\Delta}$, that correspond to $\Delta_{i1}$ or $\Delta_{i 2}$ being degenerate\footnote{We consider such $\delta\Delta$ that $\Delta(P_{m,n}) - \delta\Delta$ is not degenerate and then analytically continue the result to other~$\delta\Delta$.}. It is well known \cite{Zamolodchikov:1984rec2} that the residues at the poles are proportional to conformal blocks with the corresponding internal dimension changed to $\Delta(P_{m,-n})$. For example, for the pole at $\Delta = \Delta(P_{m,n})$ the second internal dimension of the residue block would be $\Delta(P_{m,-n})$, while in the first internal dimension we just take the limit $\Delta \to \Delta(P_{m,n})$. The blocks we get here are
\begin{equation}
\begin{aligned}
\Delta = \Delta(P_{m,n})\colon& & \quad& \mathfrak{H}^{(b)}(\{P_k\};\sqrt{P_{m,n}^2+P_{i1}^2-P_{i2}^2}, P_{m,-n}|u,q),\\
\Delta = \Delta(P_{m,n})-\delta\Delta\colon& & \quad& \mathfrak{H}^{(b)}(\{P_k\};P_{m,-n}, \sqrt{P_{m,n}^2-P_{i1}^2+P_{i2}^2}|u,q).
\end{aligned}
\end{equation}
The proportionality coefficient in the residue that corresponds to degenerate $\Delta_{i\,k}$ may be computed by considering the inverse of the corresponding Shapovalov matrix. The residue only depends on dimensions of two fields, whose OPE resulted in $V_{\Delta_{i\,k}}$ and two fields in the OPE, where $V_{\Delta_{i\,k}}$ itself is involved. So these coefficients for an $n$-point block are the same as for the $4$-point block, except for the inserted momenta. We use the results of \cite{Zamolodchikov:1984rec2, CollierRecursion} to get the following residues for the $\mathfrak{F}$ conformal block\footnote{We note that there is an alternative way to derive the residues. One could consider the analytic structure of the OPE and demand that certain unwanted poles cancel out. Then the residue of the conformal block is expressed through the ratio of structure constants in Liouville theory and the constant that comes from higher equations of motion in Liouville CFT. For details see \cite{Khromov:2025awh}.}:
\begin{equation}
\begin{aligned}
\Delta = \Delta(P_{m,n})\colon&  & R_{m,n}(P_3, \sqrt{P_{m,n}^2+P_{i1}^2-P_{i2}^2}, P_4, P_5)\, \mathfrak{F}^{(b)}(\{P_k\};\sqrt{P_{m,n}^2+P_{i1}^2-P_{i2}^2}, P_{m,-n}|x,y),\\
\Delta = \Delta(P_{m,n})-\delta\Delta\colon&  & R_{m,n}(P_1, P_2, \sqrt{P_{m,n}^2-P_{i1}^2+P_{i2}^2}, P_3)\,\mathfrak{F}^{(b)}(\{P_k\};P_{m,-n}, \sqrt{P_{m,n}^2-P_{i1}^2+P_{i2}^2}|x,y). 
\end{aligned}
\end{equation}
Here
\begin{multline}\label{eq:Recursion.R}
	R_{m,n}(P_1, P_2, P_3, P_4) 
    =
    \\
    = -\frac12 \prod_{p,q} 
	\left(P_1 + P_2 - P_{p,q}\right)
	\left(P_1 - P_2 - P_{p,q}\right)
	\left(P_3 + P_4 - P_{p,q}\right)
	\left(P_3 - P_4 - P_{p,q}\right)
	\prod_{k,l}(2P_{k,l})^{-1},
\end{multline}
and
\begin{equation}
	\begin{aligned}
		&p\in\{-m+1,-m+3, \ldots,m-3,m-1\}, \quad q\in\{-n+1,-n+3, \ldots,n-3,n-1\},\\
		&k\in\{-m+1,-m+2, \ldots,m-1,m\}, \quad l\in\{-n+1,-n+2, \ldots,n-1,n\}, \quad (k,l)\notin\{(0,0),\, (m,n)\}.
	\end{aligned}
\end{equation}
Note that despite the presence of square roots $R_{m,n}$ depend analytically on $P_{i1}$ and $P_{i2}$. For example:
\begin{multline}
R_{m,n}(P_3, \sqrt{P_{m,n}^2+P_{i1}^2-P_{i2}^2}, P_4, P_5)
=
\\
=
-\frac12 \prod_{p,q} 
	\left((P_3-P_{p,q})^2 - (P_{m,n}^2 + P_{i1}^2-P_{i2}^2)\right)
	\left(P_4 + P_5 - P_{p,q}\right)
	\left(P_4 - P_5 - P_{p,q}\right)
	\prod_{k,l}(2P_{k,l})^{-1},
\end{multline}

Finally, for the $\mathfrak{H}$ block we get an additional coordinate prefactor that corresponds to the ratio of the asymptotics $\mathfrak{f}_{\Delta}$:
\begin{equation}
\begin{aligned}
\Delta = \Delta(P_{m,n})\colon& & \quad& \frac{\mathfrak{f}_{\Delta}^{(b)}(\{P_k\};\sqrt{P_{m,n}^2+P_{i1}^2-P_{i2}^2}, P_{m,-n}|x,y)}{\mathfrak{f}_{\Delta}^{(b)}(\{P_k\};\sqrt{P_{m,n}^2+P_{i1}^2-P_{i2}^2}, P_{m,n}|x,y)} = e^{2imn(\frac{\pi}{2}\tau - u - i\log 2)},\\
\Delta = \Delta(P_{m,n})-\delta\Delta\colon& & \quad& \frac{\mathfrak{f}_{\Delta}^{(b)}(\{P_k\};P_{m,-n}, \sqrt{P_{m,n}^2-P_{i1}^2+P_{i2}^2}|x,y)}{\mathfrak{f}_{\Delta}^{(b)}(\{P_k\};P_{m,n}, \sqrt{P_{m,n}^2-P_{i1}^2+P_{i2}^2}|x,y)} = e^{2imn(u-i\log2)}.
\end{aligned}
\end{equation}
Combining all of the above and applying the Liouville theorem we get:
\begin{multline} \label{eq:rec5pblock}
\mathfrak{H}^{(b)}(\{P_k\};P_{i1}, P_{i2}|u,q) 
=
1 
+
\\
+
\sum_{m,n = 1}^{\infty} (4qe^{-2iu})^{mn}\frac{R_{m,n}(P_3, \sqrt{P_{m,n}^2+P_{i1}^2-P_{i2}^2}, P_4, P_5) }{P_{i2}^2 - P_{m,n}^2}\mathfrak{H}^{(b)}(\{P_k\};\sqrt{P_{m,n}^2+P_{i1}^2-P_{i2}^2}, P_{m,-n}|u,q)
+
\\
+
\sum_{m,n = 1}^{\infty} (4e^{2iu})^{mn}\frac{R_{m,n}(P_1, P_2, \sqrt{P_{m,n}^2-P_{i1}^2+P_{i2}^2}, P_3)}{P_{i1}^2 - P_{m,n}^2}\mathfrak{H}^{(b)}(\{P_k\};P_{m,-n}, \sqrt{P_{m,n}^2-P_{i1}^2+P_{i2}^2}|u,q).
\end{multline}

\subsubsection{Analytic checks}
The recursion we derived is valid for arbitrary conformal blocks. In particular, it is valid in the exactly solvable case $P_1 = P_2 = P_4 = P_5 = ib/4$, $P_3 = P_{1,2}$, where we should get
\begin{equation}\label{eq:ExactRecursion}
\mathfrak{H}^{(b)}\left(\frac{ib}{4},\frac{ib}{4},P_{1,2},\frac{ib}{4},\frac{ib}{4};\left.P \pm \frac{i}{2b}, P\right|u,q\right) = \left(\frac{e^{-2iu}}{4}\frac{2i\Theta_1'(0|q)}{\Theta_1(2u|q)}\right)^{-\frac{1}{4b^2}}.
\end{equation}
Note that from \eqref{eq:Useful.Theta1} it follows that if we expand the right-hand side in $q$ at each term $q^M$ the powers of $e^{iu}$ exponents would be at least $-2M$, which agrees with the form of the series we get from the recursion.  However, from the recursion it is not obvious even that this function does not depend on $P$. Since \mbox{$\delta\Delta = 2P\delta P  + \delta P^2$} depends on $P$, the $R_{m,n}$ and conformal blocks in the residues also depend on $P$. Moreover, $P_{i1}^2$ and $P_{i2}^2$ also depend on $P$, so there appear to be poles in $P$. 

One can see that in the first step of the recursion the poles are canceled by $R_{m,n}$. For terms with $P_{i2}^2$ in the denominator the $R_{m,n}$ contains the following factors
\[
\begin{aligned}
+\colon & & P_{m,n}^2 + \left(P +\frac{i}{2b}\right)^2 - P^2 - (P_{1,2} + P_{m-1,n-1})^2,\\ 
-\colon & & P_{m,n}^2 + \left(P- \frac{i}{2b}\right)^2 - P^2 - (P_{1,2} + P_{m-1,n-3})^2,
\end{aligned}
\] 
which cancel the pole at $P = P_{m,n}$ for the two choices of signs. Note that because $P_1 = P_2 = P_4 = P_5 = ib/4$, only $R_{m,n}$ with even $n$ are nonzero, hence there is always a factor with $q = n-3$. The same factors also cancel the poles for terms with $P_{i1}^2$ in the denominators.

In the second step of the recursion this simple logic fails. Consider, for example, conformal block $\mathfrak{H}^{(b)}(\{P_k\};\sqrt{P_{m,n}^2+(P+i/2b)^2-P^2}, P_{m,-n}|u,q)$ that appears in the residue in \eqref{eq:rec5pblock}. After applying the recursion formula \eqref{eq:rec5pblock} for it again consider the term with 
\[
P_{m,n}^2+\left(P+\frac{i}{2b}\right)^2-P^2 - P_{r,s}^2
\] 
in the denominator. It is multiplied by $R_{r,s}$, which is a product of
\[
P_{r,s}^2 - \left(P_{m,n}^2 + \left(P+\frac{i}{2b}\right)^2-P^2\right) + P_{m,-n}^2 - (P_{1,2} + P_{p,q})^2,
\]
so for them to cancel the pole we need 
\[
P_{1,2} + P_{p,q} = \pm P_{m,-n}.
\]
We already mentioned that $n$ is even. Since $s$ is also even, $q$ is odd. Thus, such $p$ and $q$ do not exist and such simple cancellation of the pole does not occur. Nevertheless, we checked up to terms $q^Me^{2i(N-M)u}$ with $N, M\leq 10$ that the recursion formula reproduces~\eqref{eq:ExactRecursion}.

We also checked up to terms including $y^{3}, (x/y)^{3}$ that the recursion reproduces the AGT series.

\subsubsection{Numerical checks of the elliptic recursion}

\begin{figure}[h]
    \includegraphics[width=0.55\linewidth]{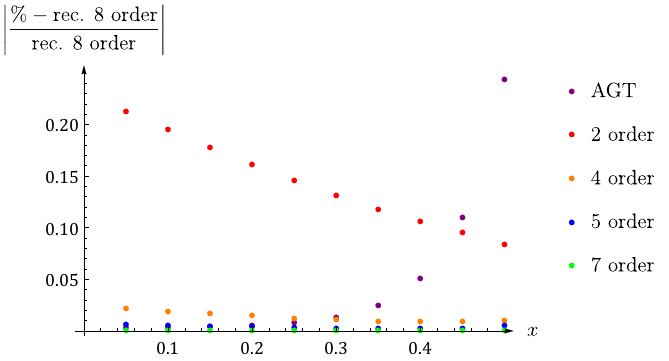}
  \includegraphics[width=0.45\linewidth]{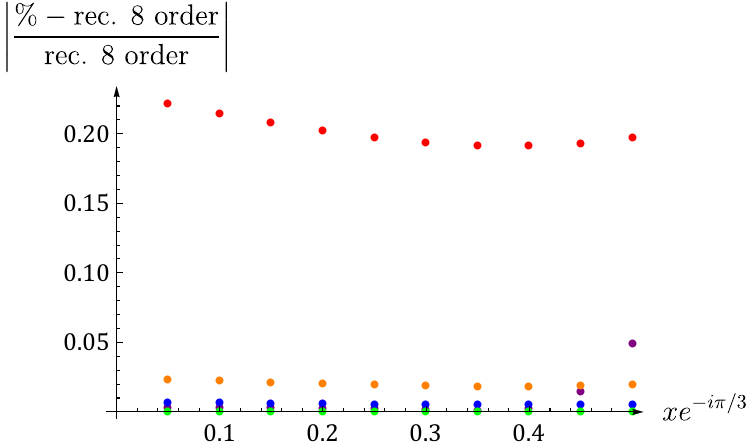}
  \caption{Relative difference between elliptic recursion at 8th order and other methods with fixed intermediate momenta $P_{i1}=1, P_{i2}=1.5$, for real $x$ (left) and $\text{Arg}\,x = \pi/3$ (right).}
    \label{fig:plotsrandblockx}
\end{figure}

\begin{figure}[h]
    \centering
    \includegraphics[width=0.65\linewidth]{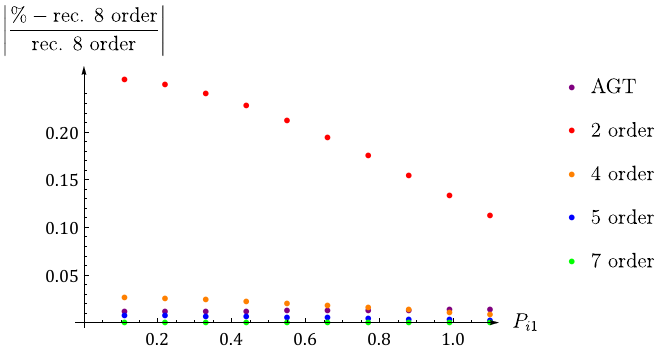}
  \caption{Relative difference between elliptic recursion for 8th order and other methods for fixed $x = 0.3$ and $P_{i2} = 1.5$ as a function of $P_{i1}$.}
    \label{fig:plotsrandblockp}
\end{figure}

\paragraph{Random block. } We compare elliptic recursion and AGT series expansion for a conformal block with ``random'' parameters:
$$
b = \frac{1}{e},\quad\lbrace P_1,P_2,P_3,P_4,P_5 \rbrace = \left\lbrace \frac{1}{3}, \frac{i}{4}, \frac{2}{7}, i \sqrt{\frac{3}{2}}, \frac{e}{\pi} \right\rbrace,\quad y=0.6.
$$
In AGT method we expand up to $10$th order in $y$ and $x/y$. In Figures \ref{fig:plotsrandblockx} and \ref{fig:plotsrandblockp}, we compare recursion truncated at different order in elliptic variables with AGT for different $x$ on the real line and some directions in the complex plane, fixing intermediate momenta $P_{i\,1,2}$, as well as for fixed $P_{i2},x$ and different $P_{i1}$. For plots in Figure \ref{fig:plotsrandblockx}, difference between AGT results and elliptic recursion quickly grows when we approach the boundary of supposed region of convergence $|x| = |y|$, while recursion behaves in a more stable way: relative difference between 7th and 8th order is already $\sim 10^{-4}$. In any case the recursion, even up to 8th order, is much faster than computing the AGT series (at least in our implementation).

\paragraph{Crossing symmetry for Liouville correlator. } As a complementary check of elliptic recursion \eqref{ell-recursion}, we use it to evaluate the probe five-point Liouville correlator and check its behaviour under some projective transformations. For transformations $z \to 1-z$ and $z \to \frac{z-x}{1-x}$ the correlator should behave as \footnote{Note that the second transformation $\frac{z-x}{1-x}$ is, in fact, an exact symmetry of our elliptic block, as can be seen from the recursion \eqref{eq:rec5pblock}. Under this transformation $\tau(x) \to \tau(x)+1,\,u(y,x) \to u(y,x) - \frac{\pi}{2}$, so $qe^{-2iu}$ is invariant and $e^{2iu}$ changes sign. The extra sign is compensated by permuting $P_1,P_2$ in $R_{m,n}(P_1,P_2,\dots)$. Thus in the cases when $|\frac{x}{1-x}|< |\frac{y-x}{1-x}|$ symmetry under this transformation is trivial. Similar statement would be valid for the transformation $z \to \frac{z}{z-1}$.}
\begin{multline}\label{eq: crossing}
\overbrace{\langle V_{P_1}(0) V_{P_2}(x) V_{P_3}(y) V_{P_4}(1) V_{P_5}(\infty) \rangle}^{(1)} = \overbrace{\langle V_{P_4}(0) V_{P_2}(1-y) V_{P_3}(1-x) V_{P_1}(1) V_{P_5}(\infty)}^{(2)} \rangle =  \\
 = |1-x|^{-2(\Delta_1 + \Delta_2 + \Delta_3 + \Delta_4 - \Delta_5)} \underbrace{\langle V_{P_2}(0) V_{P_1}\left(\frac{-x}{1-x} \right) V_{P_3}\left(\frac{y-x}{1-x} \right) V_{P_4}(1) V_{P_5}(\infty) \rangle}_{(3)} 
\end{multline}
As an example consider $\lbrace P_{1}, \dots, P_5 \rbrace = \lbrace 0.3,0.4,0.6,0.5,0.25 \rbrace$ and the central charge $b = \sqrt{\frac{2}{11}}$. We use normalization conventions and formula for structure constants of \cite{Collier:2023cyw}. For such central charge, Liouville structure constants can be expressed in terms of Barnes G-function (again, see \cite{Collier:2023cyw}), which has a standard implementation in Wolfram Mathematica and can be computed efficiently. The integral over intermediate momenta is calculated by approximating it with the Riemann sum with step $0.05$ for each momenta. In Table \ref{tab:crossing}, numerical results for correlators $(1),(2),(3)$ for different values of $(x,y)$ are presented. 
\begin{table}
    \centering
    \begin{tabular}{|c|c|c|c|c|c|}
    \hline
        $(x,y)$ & rec.order & $(1)$ & $(2)$ & $(3)$ & $|1-x|^{-2\left(\sum \Delta_i-2\Delta_5\right)}$ \\
        \hline
        $(0.3,0.7)$ & 5 & $3.494\cdot10^{-8}$ & $3.495\cdot 10^{-8}$ & $3.246\cdot 10^{-10}$ & $107.64$\\
         \hline
         $(0.35,0.85)$ & 6 & $1.282\cdot10^{-7}$ & $1.291\cdot 10^{-7}$ & $4.507\cdot 10^{-10}$ & $284.55$\\
         \hline
         $(0.3,0.4)$ & 6 & $6.565\cdot 10^{-7}$  & $6.549\cdot 10^{-7}$ & $6.102 \cdot 10^{-9}$  & $107.64$ \\
         \hline
         $(0.2(1+i),0.7)$ & 5& $1.644 \cdot 10^{-8}$ & $1.652 \cdot 10^{-8}$ &  $1.31 \cdot 10^{-9}$ & $12.55$ \\
         \hline
    \end{tabular}
    \caption{Checks of crossing symmetry. $(1),(2)$ and $(3)$ denote values of correlators in \eqref{eq: crossing}.}
    \label{tab:crossing}
\end{table}
While accuracy somewhat decreases when $x \to y$ or $y \to 1$, we see that crossing symmetry can be established with good precision, using reasonable recursion depth.

\subsection{$\Delta$-recursion for $N$-point blocks}
Similarly to what we did for the $5$-point case, we now use the asymptotics of the $N$-point block to write a recursion relation. To this end we define the elliptic block:
\begin{equation}
\mathfrak{F}^{(b)}(\{P_k\}; \{P_{i\,k}\}|x,\{y_k\}) 
=
\mathfrak{f}^{(b)}_{\Delta}(\{P_k\}; \{P_{i\,k}\}|x,\{y_k\})\mathfrak{H}^{(b)}_{\Delta}(\{P_k\}; \{P_{i\,k}\}|\{u_k\},q).
\end{equation}
Here we again denote $u_k \equiv u(y_k,x)$. We then have
\begin{equation}
\lim_{\substack{P_{i1}, \ldots, P_{i\,N-3}\to \infty\\ P_{i\,k}^2 - P_{i\,k+1}^2~\text{fixed}}} \mathfrak{H}^{(b)}_{\Delta}(\{P_k\}; \{P_{i\,k}\}|\{u_k\},q)
=
1.
\end{equation}
The $\mathfrak{H}$ block with internal dimensions $\Delta_{i\,N-3} = \Delta$, $\Delta_{i,\,N-4} = \Delta + \delta\Delta_{N-4}$,~$\ldots$, $\Delta_{i,1} = \Delta + \sum_{l=1}^{N-4}\delta\Delta_l$ has a regular point at infinity as a function of $\Delta$ (with $\delta\Delta_l = P_{i\,l}^2-P_{i\,l+1}^2$ fixed). The poles of $\mathfrak{H}$ form $N-3$ series, where each series corresponds to some internal dimension being degenerate. We denote the conformal block that appears in the residue when $\Delta = \Delta(P_{m,n})-\sum_{l=k}^{N-4}\delta\Delta_l$, i.\,e. when $\Delta_{i\, k}$ is degenerate, as $\mathfrak{H}^{m,n}_k(\{P_l\}, \{P_{i\,l}\}|\{u_l\},q) $. It has the following internal momenta:
\begin{equation}
\begin{aligned}
1\leq l<k:& &\qquad &P_{i\, l} = \sqrt{P_{m,n}^2 + \sum_{j=l}^{k-1}\delta\Delta_j},\\
l=k:& & &P_{i\,l} = P_{m,-n},\\
N-3\geq l>k:& & &P_{i\,l} = \sqrt{P_{m,n}^2 - \sum_{j=k}^{l-1}\delta\Delta_j}.
\end{aligned}
\end{equation}
The corresponding ratio of the asymptotics is
\begin{equation}
\begin{aligned}
k = 1:& &\qquad & e^{2imn(u_1- i\log 2)},\\
k = 2, \ldots, N-4:& & &e^{2imn(u_k - u_{k+1})},\\
k = N-3:& & &e^{-2imn (u_{N-4}-\frac{\pi}{2}\tau + i\log 2)}.
\end{aligned}
\end{equation}
Then, applying the Liouville theorem we get
\begin{multline}
\mathfrak{H}^{(b)}_{\Delta}(\{P_k\}; \{P_{i\,k}\}|\{u_k\},q)
=
1 
+
\\
+
\sum_{m,n = 1}^{\infty} e^{2imn (\frac{\pi}{2}\tau - u_{N-4} - i\log 2)}\frac{R_{m,n}(P_{N-2}, \sqrt{P_{m,n}^2+\delta\Delta_{N-4}}, P_{N-1}, P_N)} {P_{i\,N-3}^2 - P_{m,n}^2}\mathfrak{H}^{m,n}_{N-3}(\{P_l\}, \{P_{i\,l}\}|\{u_l\},q) 
+
\\
+
\sum_{k=2}^{N-4} \sum_{m,n = 1}^{\infty} e^{2imn (u_k - u_{k+1})}\frac{R_{m,n}(\sqrt{P_{m,n}^2+\delta\Delta_{k-1}}, P_{k+1}, \sqrt{P_{m,n}^2-\delta\Delta_k}, P_{k+2}) }{P_{i\, k}^2 - P_{m,n}^2}
\mathfrak{H}^{m,n}_k(\{P_l\}, \{P_{i\,l}\}|\{u_l\},q) 
+
\\
+
\sum_{m,n = 1}^{\infty} e^{2imn (u_1-i\log 2)}\frac{R_{m,n}(P_1, P_2, \sqrt{P_{m,n}^2-\delta\Delta_1}, P_3)}{P_{i1}^2 - P_{m,n}^2}\mathfrak{H}^{m,n}_1(\{P_l\}, \{P_{i\,l}\}|\{u_l\},q) .
\end{multline}
Here, as in the $5$-point case, we used the results of \cite{CollierRecursion} to write the coefficients $R_{m,n}$ in the residues.

\subsection{Amplitudes in Liouville gravity}
\subsubsection{Setting up the problem}

As mentioned in the introduction, we consider string theories with the worldsheet CFT consisting of 3 sectors:  $BC$-system of BRST-ghosts, Liouville CFT and some other CFT as ``matter''. We will concentrate on Virasoro minimal model as matter (``minimal string theory'').  We will denote kinematical quantities (like the central charge, dimensions, etc.) related to the matter sector by a hat, e.\,g. $\hat{c},\,\hat{b},\,\hat{\Delta}(\hat{P})$. A solution for the central charge constraint ($\hat{c} + c = 26$) is $\hat{b} = ib$. Also the primary fields in the matter sector will be denoted as $\Phi_{\hat{P}}$ to distinguish them from Liouville primaries $V_P$.

We will only recap general information relevant to define what we are computing. For more thorough introduction see other works, e.\,g. \cite{Belavin:2005jy}.

The worldsheet theory has a BRST-charge, the cohomology of which defines physical operators. Their spectrum depends on the matter theory and in general is nonempty for all ghost numbers. We will consider two types of operators in this work:
\begin{enumerate}
\item the ``tachyon'' operators of ghost number $2$, built by combining primary fields from Liouville and matter sector such that their total conformal dimension is $(1,1)$ and dressing additionally with $C$-ghosts:
$$
\mathcal{T}_P = C \br{C} V_P \Phi_{iP}.
$$
We have solved the dimension constraint explicitly: $\Delta(P) + \hat{\Delta}(iP) = 1$. Such operators exist for all studied choices of matter theory and have been examined the most thoroughly.
\item the ``ground ring'' operators of ghost number zero exist if the matter theory has degenerate primary fields $\Phi_{m,n}$. Ground ring operators have the general form 
$$
O_{m,n} = H_{m,n} \br{H}_{m,n} V_{m,n} \Phi_{m,n}
$$
where $H_{m,n} = L_{-1}^{mn} + \dots$ is a polynomial of degree $mn-1$ built from Virasoro modes $L_{-k}, M_{-k}$ (in Liouville and matter sector, respectively) and ghost modes $B_{k},C_{k}$, acting on $V\Phi$. No general formula is known, but in specific examples requirement of BRST-closedness allows to find them explicitly \cite{imbimbo1992}; the main example is
$$
H_{2,1} = L_{-1} - M_{-1} + b^2 B_{-2} C_1.
$$
Interpreted as a field, this ground ring operator 
\begin{equation}
O_{2,1}(z) = (\pd_L - \pd_M + b^2 \colon\!BC\colon\!(z)) \br{(\dots)} V_{2,1} \Phi_{2,1}(z),
\end{equation}
where $\pd_{L/M}$ are derivatives that act only on Liouville/matter field in the product. 
\end{enumerate}
Correlators of $n$ physical operators on some surface of genus $g$ give rise to closed differential forms on the moduli space of complex curves $\mathcal{M}_{g,n}$, which can then be integrated to obtain string amplitudes. The general prescription to construct the integrand is given e.\,g. in \cite{polchinski_1998, Sen:2024nfd}. Degree of the form depends on the total ghost number: if we take $n$ tachyon operators, the result of this construction is a top degree ($6g-6+2n$)-form, so it gives a number after integrating over all moduli space. For example, in case of genus $0$ the amplitude can be computed as the integral over $(n-3)$ cross-ratios that parametrize $\mathcal{M}_{0,n}$
\begin{equation}
\int \prod \limits_{i=1}^{n-3} d^2z_i \left\langle \mathcal{T}_{P_{n-2}}(0) \mathcal{T}_{P_{n-1}}(1) \mathcal{T}_{P_{n}}(\infty)  \prod \limits_{i=1}^{n-3} V_{P_i} \Phi_{iP_i}(z_i) \right\rangle.
\end{equation}
However, if we try to incorporate $k$ ground ring operators in the construction, degree of the form drops by $2k$. On the sphere, correlator of $n-3$ ground ring operators and $3$ tachyons is well-defined, being a closed 0-form on $\mathcal{M}_{0,n}$, i.\,e. a number. But, for example, correlator of 1 ground ring operator $O_{2,1}$ and 4 tachyons on $\mathcal{M}_{0,5}$ gives a ``ground ring 2-form'':
\begin{align*}
 &\langle  \mathcal{T}_{P_{1}}(0) O_{2,1}(y) \left(V_{P_2} \Phi_{iP_2}\right)(x) \mathcal{T}_{P_{4}}(1) \mathcal{T}_{P_{5}}(\infty)  \rangle dx\wedge d\br{x} - \\
 &- b^2 \langle  \mathcal{T}_{P_{1}}(0) B\br{H}_{2,1} V_{2,1} \Phi_{2,1}(y) \left(CV_{P_2} \Phi_{iP_2}\right)(x) \mathcal{T}_{P_{4}}(1) \mathcal{T}_{P_{5}}(\infty)  \rangle dy\wedge d\br{x} - \\
 &- b^2 \langle  \mathcal{T}_{P_{1}}(0) \br{B}H_{2,1} V_{2,1} \Phi_{2,1}(y) \left(\br{C} V_{P_2} \Phi_{iP_2}\right)(x) \mathcal{T}_{P_{4}}(1) \mathcal{T}_{P_{5}}(\infty)  \rangle dx \wedge d\br{y} - \\
 &-b^4  \langle  \mathcal{T}_{P_{1}}(0) B\br{B} V_{2,1} \Phi_{2,1}(y) \mathcal{T}_{P_2}(x) \mathcal{T}_{P_{4}}(1) \mathcal{T}_{P_{5}}(\infty)  \rangle dy\wedge d\br{y}.
\end{align*}
The fact that the form is closed can be understood using the identities for BRST-descendants
\begin{align*}
    \pd H_{2,1} V_{2,1}\Phi_{2,1} &= \mathcal{Q} \left(b^2 B V_{2,1} \Phi_{2,1} \right), \\
    \pd (C V_P \Phi_{iP}) &= \mathcal{Q} \left(V_P \Phi_{iP} \right),
\end{align*}
and moving the action of BRST-charge from one operator to the other inside the correlator. Equivalently, one can just use conformal Ward identities and BPZ equation. After choosing a 2-cycle, we can integrate a 2-form over it (e.\,g. over $(y,\br{y})$ at fixed $x$ and vice versa) to obtain a ``string amplitude with ground ring''. 

In the next subsection, we will test the applicability of our expressions for 5-point conformal blocks in numerical studies of such amplitudes. The reason we introduce ground ring in the picture is that, while still probing the higher-point kinematics, such amplitudes are expected to be easier to compute than purely tachyonic ones. Namely, they involve only integrals over simple submanifolds of $\mathcal{M}_{0,5}$ and due to the presence of degenerate fields CFT correlators that we need to compute have a simpler structure than a general 5-point function. Another motivation is to test a recent proposal about how ground ring operators (or something closely related) can emerge in ``matrix model'' dual for the minimal string \cite{Artemev:2025pvk}. 
\subsubsection{Computation of the integral over $d^2x$}
\begin{figure}[h!]
 	\begin{subfigure}[b]{0.49\linewidth}
 		\centering
        \includegraphics[width=.95\linewidth]{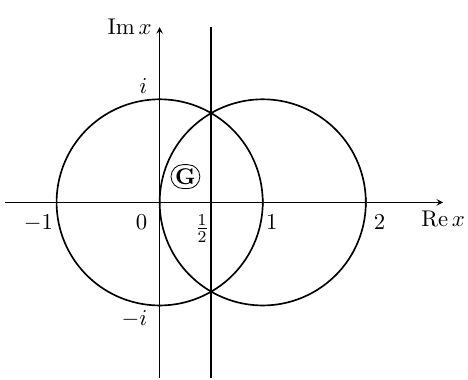}
 		\caption{Images of the fundamental domain $\mathbf{G}$.}\label{fig:domainsz}
 	\end{subfigure}~
 	\begin{subfigure}[b]{0.49\linewidth}
 		\centering
        \includegraphics[width=.75\linewidth]{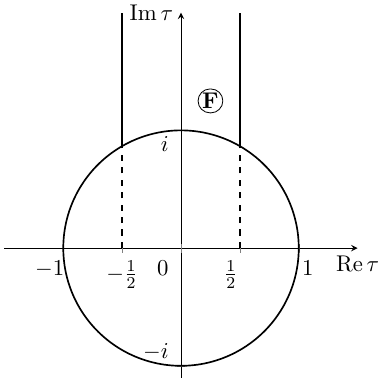}
 		\caption{Image $\mathbf{F}$ of the domain $\mathbf{G}$.}\label{domainstau}
 	\end{subfigure}
 	\caption{}\label{fig:domains}
 \end{figure}

Here we illustrate how one can use elliptic recursion of section \ref{ell-recursion} for numerical computations of the integral of ``ground ring 2-form'' over the position $x$ of one of the tachyons. We consider as an example the case of $(2,9)$ minimal model ($b = \sqrt{2/9}$) as the matter theory, with external tachyon momenta
\begin{equation}
   \lbrace P_1, P_2, P_4, P_5 \rbrace = \lbrace P_{-3,1}, P_{-2,1}, P_{-3,1}, P_{-3,1} \rbrace.
\end{equation}
These Liouville fields ``dress'' minimal model primaries $\Phi_{3,1}, \Phi_{2,1}, \Phi_{3,1}, \Phi_{3,1}$ respectively. Amplitude is of the form
\begin{align*}
A=   \int d^2 x\, \bar{\hat{H}}_{2,1}\hat{H}_{2,1}\langle 
\Phi_{3,1}(0)\Phi_{2,1}(x) \Phi_{2,1}(y)\, \Phi_{3,1}
(1)\,\Phi_{3,1} (\infty) \rangle \times \\
\times \langle 
V_{-2,1}(0)V_{-2,1}(x) V_{2,1}(y)\, V_{-3,1}
(1)\,V_{-3,1} (\infty) \rangle 
\end{align*}
where $\hat{H}_{2,1} = \pd_y^L - \pd_y^M + b^2 \frac{2y-1}{y(1-y)}$, because the ghost correlator is
\begin{align*} 
\langle C(0) B(y) C(x) C(1) C(\infty) \rangle &= \frac{x(1-x)}{y(1-y)(y-x)}, \\
\langle C(0) \colon\! BC\colon\!(y) C(1) C(\infty) \rangle &= \frac{2y-1}{y(1-y)}.
\end{align*}
A subgroup $\mathcal{S}$ of projective transformations of order $6$ generated by $\mathcal{A}$: $z\rightarrow 1/z$ and $\mathcal{B}$: $z\rightarrow
1-z$ can be used to divide the $x$ plane into 6 regions --- images of the fundamental domain $\mathbf{G}=\{\operatorname*{Re}x<1/2;\;\left| 1-x\right| <1\}$ (see Fig.~\ref{fig:domainsz}).
 
$O_{2,1}$ and $\mathcal{T}_{k,1}$ are primary operators of dimension zero, so the correlator transforms in a simple way under the action of $\mathcal{S}$ and we can reduce the integration over $d^2x$ to 6 integrals over~$\mathbf{G}$:
\begin{align}
A = a\left(y\right)+a\left(\frac{1}{y}\right)+a\left(1-y\right)+a\left(\frac{y-1}{y}\right)+a\left(\frac{1}{1-y}\right)+a\left(\frac{y}{y-1}\right),\, \nonumber\\
a(y) := \int\limits_{\mathbf{G}}d^2x\, \langle \mathcal{T}_{P_{1}}(0)  \left(V_{P_2} \Phi_{iP_2}\right)(x) O_{2,1}(y)\mathcal{T}_{P_{3}}(1) \mathcal{T}_{P_{4}}(\infty) \rangle .
\end{align}
Finally, in every term one can decompose the correlators in conformal blocks. E.\,g. let us expand in the comb channel $(0,x,y,1,\infty)$; the contribution to the amplitude then takes the form
\begin{equation}
a(y) = \sum \limits_{(kl)} r_{kl} \sum_{\epsilon=\pm}\int
\frac{dP_{i1}}{4\pi} r_{\epsilon}(P_{i1}) \underbrace{\int \limits_\mathbf{G} d^2x\, |\hat{H}_{2,1} \mathfrak{F}^{(b)}(x,y) \mathfrak{F}^{(ib)}(x,y)|^2}_{I_{\epsilon, (kl)}(P_{i1},y)},
\end{equation}
where we omit most of the parameters of the blocks for brevity:
\begin{align*}
\mathfrak{F}^{(b)}(x,y) &:= \mathfrak{F}^{(b)}(P_1, P_2, P_{2,1},P_4,P_5;P_{i1},P_{i2}|x,y),   \\
 \mathfrak{F}^{(ib)} (x,y) &:= \mathfrak{F}^{(ib)}(iP_1, iP_2, \hat{P}_{2,1},iP_4,iP_5;\hat{P}_{i1},\hat{P}_{i2}|x,y)   . 
\end{align*}
$r_{kl}$ and $r_\epsilon (P_{i1})$ are products of structure constants in matter and Liouville theory, respectively. They are indexed by intermediate channels $(k,l)$ on the minimal model side such that $\hat{P}_{i1} = \hat{P}_{k,1},\,\hat{P}_{i2} = \hat{P}_{l,1}$ and a sign $\epsilon = \pm$ on Liouville side such that $P_{i2} = P_{i1} + \epsilon \frac{ib}{2}$. In our example $(k,l)$ take values $(4,5), (4,3), (2,3), (2,1)$.

For demonstrative purposes we will concentrate on studying the integral $I_{\epsilon,(kl)} (P,y)$ over moduli space as a function of intermediate dimensions. We will take $y=1/2$: this choice is the most symmetric, since its orbit under $\mathcal{S}$ consists only of 3 different points instead of 6. Also, it is easy to check that two choices for $\epsilon = \pm$ differ by complex conjugation, so we take $\epsilon = +$. Computing its values on a fine enough lattice in $P$-space, one can multiply it by structure constants, which are smooth functions of $P_{i1}$ and approximate the remaining integral by the Riemann sum. Numerical computation of structure constants is a simpler problem, where many effective approaches were developed, such as the infinite product formula \cite{clessthan12015} and the aforementioned expression as a finite product of Barnes G-functions for rational $b^2$.

Under the elliptic map $x \to \tau(x) = i \frac{K(1-x)}{K(x)}$ region $\mathbf{G}$ maps $1:1$ to a standard ``keyhole'' region $\mathbf{F}$ on $\tau$ upper half-plane $\lbrace \text{Im\,}\tau>0, |\text{Re\,}\tau|<1/2, |\tau|>1 \rbrace$ (see Fig.~\ref{domainstau}), which is convenient to integrate over when using elliptic variables. 

\begin{figure}[h]
    \includegraphics[width=0.45\linewidth]{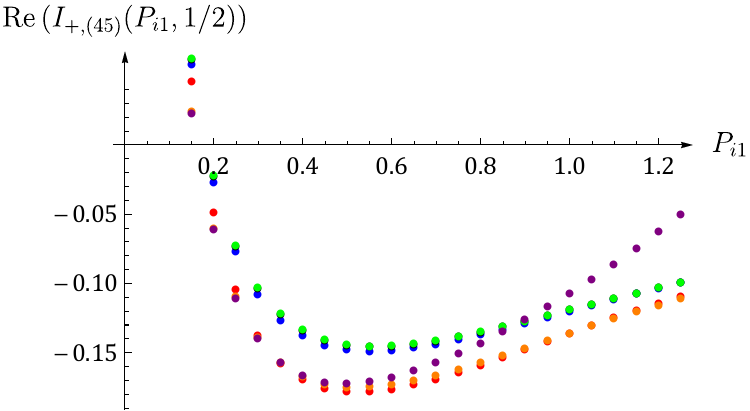}
  \includegraphics[width=0.55\linewidth]{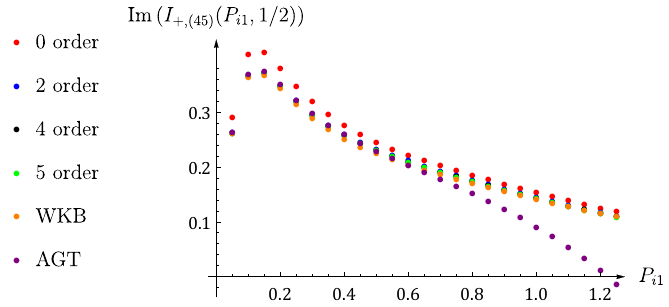}
  \caption{Real and imaginary part of $I_{+,(45)}(P_{i1},1/2)$ computed using different approaches.}
    \label{fig:45plots}
\end{figure}

\begin{figure}[h]
    \includegraphics[width=0.5\linewidth]{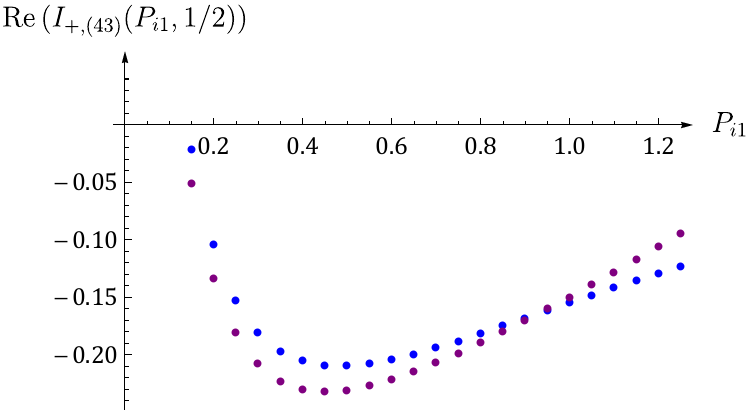}
  \includegraphics[width=0.5\linewidth]{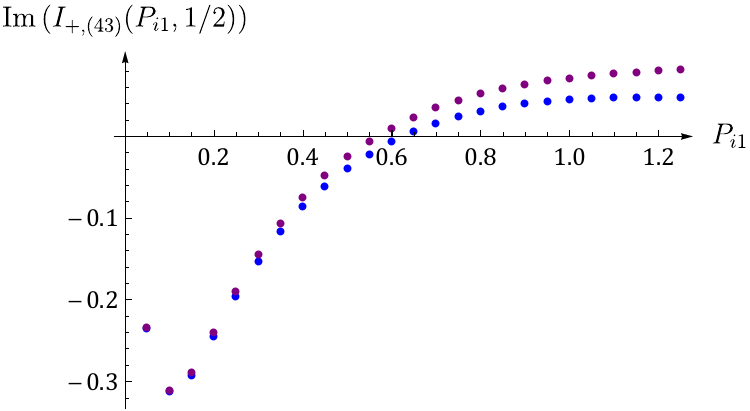}
      \includegraphics[width=0.5\linewidth]{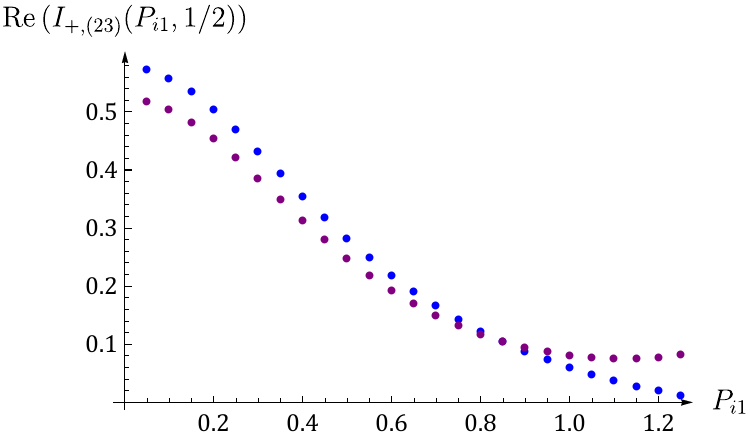}
  \includegraphics[width=0.5\linewidth]{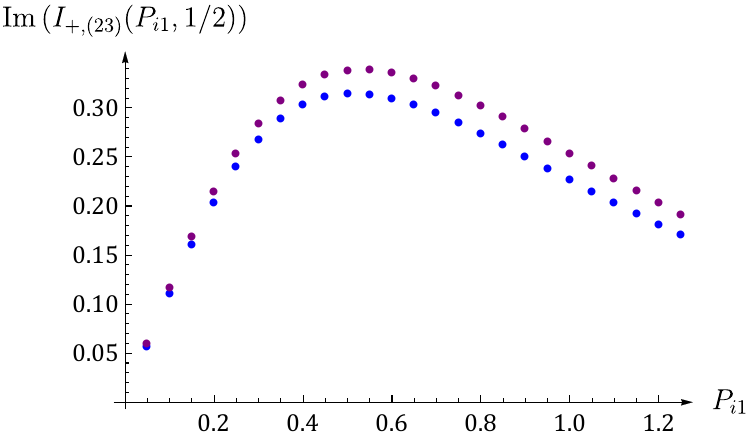}
      \includegraphics[width=0.5\linewidth]{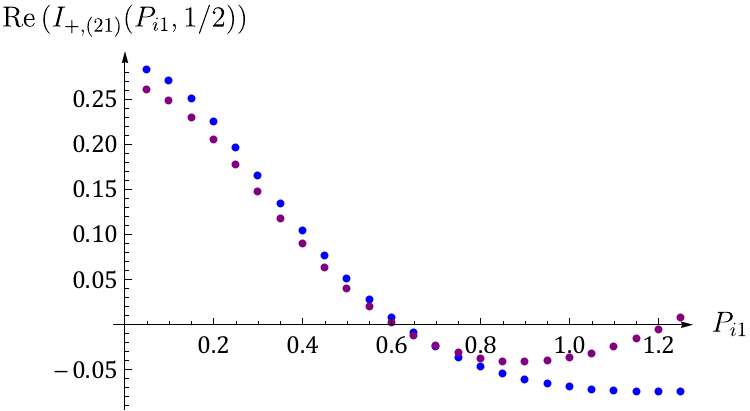}
  \includegraphics[width=0.5\linewidth]{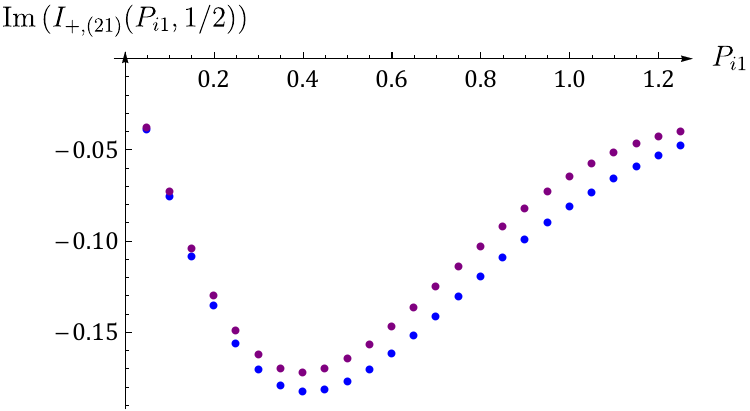}
  \caption{Real and imaginary part of $I_{+,(kl)}(P_{i1},1/2)$ for different $(kl)$ computed using recursion to 5th order (blue) and AGT expansion (purple).}
    \label{fig:otherplots}
\end{figure}

The result of action of $\hat{H}_{2,1}$ on a product of conformal blocks given by \eqref{elliptic-5ptblock-def} simplifies to
\begin{align}
dx\,&H_{2,1} \mathfrak{F}^{(b)}(x,y) \mathfrak{F}^{(ib)}(x,y) = d\tau\,\pi\, 16^{P_{i1}^2 + \hat{P}_{i1}^2} q^{P_{i2}^2 + \hat{P}_{i2}^2} e^{2i (u+i\log 2) (P_{i1}^2 + \hat{P}_{i1}^2 - P_{i2}^2 - \hat{P}_{i2}^2)} \times \nonumber\\
& \times \left[ \frac{i}{2} \left(\mathfrak{H}^{(ib)}(u,q)\pd_u \mathfrak{H}^{(b)}(u,q) - \mathfrak{H}^{(b)}(u,q)\pd_u \mathfrak{H}^{(ib)}(u,q) \right) + \right.\nonumber \\
& \left. +\left(\frac{b^2 \theta_3(q)^2}{4} \sqrt{y(1-y)(y-x)} \left(\frac{3}{y-x} + \frac{2y-1}{y(1-y)} \right) \right) \mathfrak{H}^{(b)}(u,q)\mathfrak{H}^{(ib)}(u,q) - \right.\nonumber \\
& \left. - \left(P_{i1}^2 -P_{i2}^2 + \hat{P}_{i2}^2 - \hat{P}_{i1}^2 \right) \mathfrak{H}^{(b)}(u,q)\mathfrak{H}^{(ib)}(u,q) \right],
\end{align}
where we used that
\begin{equation*}
dx=\pi x(1-x)\theta_3^4(q)d\tau\;.
\end{equation*}
Interesting simplification occurs also if we act with $H_{2,1}$ on WKB answer \eqref{eq:asymptotics} that accounts for $\delta P^2$ corrections; in the considered case $\delta P^2 = - \delta \hat{P}^2 = - \frac{b^2}{4}$, so one gets
\begin{multline}
 dx\,H_{2,1} \mathfrak{f}^{(b)}(x,y) \mathfrak{f}^{(ib)}(x,y)   = d\tau \,\frac{i\pi}{2} (16q)^{P_{i2}^2 + \hat{P}_{i2}^2} e^{4(iu + \log 2) (P_{i2} \delta P + \hat{P}_{i2} \delta \hat{P})} \times  \\
  \times \left(2b^2 \frac{\Theta_1'(u,q)}{\Theta_1(u,q)} + 4i (P_{i2} \delta P - \hat{P}_{i2} \delta \hat{P}) \right)
\end{multline}
Prefactors dependent on external dimensions almost cancel the Jacobian $x\to \tau$. As an aside, a similar simplified expression exists for a different component of the ``ground ring 2-form'', which we do not consider in this paper:
\begin{multline}
dy\, \langle C(0) B(y) C(x) C(1) C(\infty) \rangle \mathfrak{f}^{(b)}(x,y) \mathfrak{f}^{(ib)}(x,y) = \\
=dy\, \langle C(0) B(y) C(x) C(1) C(\infty) \rangle \mathfrak{f}^{(b)}_\Delta(x,y) \mathfrak{f}^{(ib)}_\Delta(x,y) =  \\
= du\,(-2i) (16q)^{P_{i2}^2 + \hat{P}_{i2}^2} e^{4(iu + \log 2) (P_{i2} \delta P + \hat{P}_{i2} \delta \hat{P})}
\end{multline}
Now we present results of the computation. The plots on Figures \ref{fig:45plots}, \ref{fig:otherplots} illustrate $I_{+,(kl)}(P,1/2)$ calculated using recursion \eqref{ell-recursion}, truncated to some level in elliptic variables $qe^{-2iu}$ and $e^{2iu}$, WKB asymptotic \eqref{eq:asymptotics} and AGT series expansion\footnote{We use AGT series expansion as follows: we compute numerically the blocks and their derivatives as series in $x/y$ and $y$ up to order $n=9$ in both variables (for higher orders computation time quickly grows and numerical mistakes are accumulated), then rewrite blocks as a series in $q$ and $y$ and integrate their product over $d^2\tau$. We only expand the integrand up to $q^{\lfloor n/2 \rfloor} =q^4$, because the coefficients of the series in $q$ are themselves series in $y$ and we need enough terms for these coefficients to converge. This method is expected to have low accuracy.}. In Table \ref{tab:integrals}, results of the calculation for a few values of $P_{i1}$ using different methods are provided. We compare all available methods for the channel $(k,l)=(4,5)$, which is expected to give the largest contribution to $a(1/2)$. Two subtle points about this computation:
\begin{itemize}
    \item to compute integral over $d^2\tau$ we use standard numerical integration method in Wolfram Mathematica for a subdomain of $\mathbf{F}$ $\lbrace 0<\text{Im\,}\tau<\Lambda, |\text{Re\,}\tau|<1/2, |\tau|>1 \rbrace$ with sufficiently large $\Lambda$. The integral over the remaining part with $\text{Im\,}\tau>\Lambda$ is taken by truncating to the leading power in $q$-expansion of the integrand and integrating this analytically. This is enough for desired accuracy, as can be checked by varying $\Lambda$ and checking that the sum of two contributions remains the same.
    \item Since function $u(y,x)$ is multivalued in the region $x \in \mathbf{G}$, it is important to check that its values on the correct sheet are used in the calculation. Correct branch of $u(y,x)$ is defined by the requirement $\br{u(y,x)} = -u(\br{y}, \br{x})$.
\end{itemize}

We see that recursion method stabilizes quickly: results for 4th and 5th order are almost indistinguishable, and second order is quite close (difference of order $2\%$). Imaginary part of WKB answer agrees well with recursion, but the real part agrees better with recursion at 0 order and AGT answers at small momenta. Discrepancy between WKB and high order recursion is of order $10-15\%$. AGT results start to deviate from other methods significantly at $P_{i1} \sim 0.8$. 

\begin{table}[]
\small
\begin{tabular}{|l|l|l|l|l|l|}
\hline
$P$  & 0.2 & 0.4 & 0.6 & 0.8 & 1.0 \\ \hline
\multicolumn{6}{|c|}{$(k,l) = (4,5)$} \\ \hline
WKB \eqref{eq:asymptotics}  & $ -0.060 + 0.343 i$ & $ -0.167 + 0.251 i$  & $ -0.173 + 0.203 i$ & $ -0.157 + 0.170 i$  & $ -0.136 + 0.141 i$ \\ \hline
rec. 2nd order & $-0.027 + 0.348 i$ & $-0.138 + 0.259 i$  & $-0.148 + 0.211i$  & $-0.137 + 0.176 i$  & $-0.120 + 0.145 i$ \\ \hline
rec. 4nd order & $-0.023 + 0.347 i$& $-0.134 + 0.257 i$& $-0.145 + 
 0.209 i$& $-0.135 + 
 0.174 i$&$ -0.119 + 0.143 i$\\ \hline
rec. 5th order & $-0.022 + 0.346 i$& $-0.133 + 0.257 i$& $-0.145 + 0.209 i$& $-0.135 + 0.173i$& $-0.119 + 0.143 i$\\ \hline
\multicolumn{6}{|c|}{$(k,l) = (4,3)$} \\ \hline
rec. 4nd order & $-0.105 - 0.245 i$ & $ -0.206 - 0.086 i$ & $ -0.204 - 
 0.006 i$ & $ -0.182 + 0.031 i$ & $ -0.155 + 0.046 i$\\ \hline
rec. 5th order & $-0.104 - 0.245 i$ & $ -0.205 - 0.086 i$ & $ -0.204 - 
 0.006 i$ & $ -0.181 + 0.031 i$ & $ -0.154 + 0.046 i$\\ \hline
\multicolumn{6}{|c|}{$(k,l) = (2,3)$} \\ \hline
rec. 4nd order & $0.504 + 0.204 i$ & $ 0.354 + 0.304 i$ & $ 0.219 + 
 0.311 i$ & $ 0.123 + 0.275 i$ & $ 0.060 + 0.227 i$\\ \hline
rec. 5th order & $0.505 + 0.204 i$ & $ 0.355 + 0.303 i$ & $ 0.219 + 
 0.310 i$ & $ 0.123 + 0.274 i$ & $ 0.060 + 0.227 i$\\ \hline
\multicolumn{6}{|c|}{$(k,l) = (2,1)$} \\ \hline
rec. 4nd order & $0.225 - 0.135 i$ & $ 0.105 - 0.182 i$ & $ 0.008 - 
 0.161 i$ & $ -0.046 - 0.120i$ & $ -0.069 - 0.081 i$
\\ \hline
rec. 5th order & $0.226 - 0.135 i$ & $ 0.105 - 0.182 i$ & $ 0.008 - 
 0.161 i$ & $ -0.046 - 0.120 i$ & $ -0.069 - 0.081 i$\\ \hline
\end{tabular}
\caption{Value of the integrals $I_{+,(kl)}(P_{i1},1/2)$ for different minimal model intermediate momenta and $P_{i2} = P_{i1} + \frac{ib}{2}$.}
\label{tab:integrals}
\end{table}

\section{Conclusion}\label{sec5}
We finish by listing some possible directions for future work:
\begin{itemize}
    \item While higher genus blocks are likely to be untractable using this method, it is interesting to study the case of spherical blocks in other channels (for five-point block only comb channel blocks exist, but starting with 6 points, other options appear, such as the trifundamental channel). It is tempting to conjecture that the geometric interpretation given in section \ref{sec3.5} still holds and conformal blocks in the limit of large intermediate dimensions are computed using the period matrix of the WKB curve. To this end one can investigate how the large-$P$ limit of the crossing matrix that relates blocks in different channels agrees with this proposal: it is known that e.\,g. the modular transformation in this limit is simply a Fourier transform \cite{Nemkov:2013qma}. 
    \item After successfully testing our elliptic recursion in the context of Liouville gravity, it would be nice to finish the calculation of the amplitude with ground ring. To compute other contributions to the amplitude (such as $a(2)$), one needs to go away from the standard comb channel kinematic domain $0<|x|<|y|<1$. It would be most convenient to just analytically continue the blocks outside of this domain without changing the expansion channel, using our elliptic representation, but in practice elliptic functions such as $u(y, x)$ behave in a very weird way when evaluated e.\,g. for $|y|>1$. This issue need to be addressed. It is also interesting to apply elliptic blocks for more complicated cases, such as 5-point tachyon amplitudes in MS or VMS.
    \item For four-point amplitudes elliptic recursion allows to prove important analyticity property of the amplitudes in MS and VMS \cite{Artemev:2025pvk} --- lack of discontinuities associated to ``descendant on-shell poles''. This is an important part of ``analytic bootstrap'' argument presented in \cite{Khromov:2025awh} that fixes the possible expression for four-point amplitude in VMS. Whether the same can be proven for higher-point amplitudes remains to be understood. Perhaps the elliptic recursion will be useful in this context as well.
\end{itemize}

\section*{Acknowledgements}
We are grateful to Alexey Litvinov, Andrei Marshakov and Vladimir Belavin for fruitful discussions on this and related topics and interest in this work. This work was supported by Basis foundation. The work performed in Landau Institute has been supported by the Russian Science Foundation under the grant 23-12-00333.
\appendix

\section{Useful formulas} \label{app:formulas}
An alternative definition of $u(y,x)$:
\begin{equation}
u(y,x) = \frac{\pi}{2} - \frac{\pi}{4K(x)}\int_{0}^{\frac{y}{x}}\frac{dt}{\sqrt{t(1-t)(1-xt)}}.
\end{equation}
Derivatives of $u(y,x)$:
\begin{equation} \label{uderiv}
\frac{\partial u(y,x)}{\partial y} = \frac{i\pi}{4\sqrt{y(1-y)(y-x)} K(x)}, \quad \frac{\partial u(y,x)}{\partial x} = \frac{\pi}{4i} \frac{(1-y)(xK(x) + (y-x)\Pi(y,x))}{x(1-x)\sqrt{y(1-y)(y-x)}K(x)^2}.
\end{equation}
Some properties of theta functions:
\begin{equation}\label{eq:Useful.Theta1}
\Theta_1(u|q) = 2 \sum_{n=0}^\infty (-1)^n q^{\left(n+\frac12\right)^2}\sin((2n+1)u).
\end{equation}
\begin{equation}
\Theta_1(u+\pi(m+n\tau)|q) = (-1)^{m+n}q^{-n^2}e^{-2inu}\Theta_1(u|q).
\end{equation}
\begin{equation}
\Theta_1(2u|q) = \frac{2\Theta_1(u|q)\Theta_2(u|q)\Theta_3(u|q)\Theta_4(u|q)}{\Theta_1'(0|q)}.
\end{equation} 
\begin{equation}
\Theta'_1(0|q) = \theta_2(q)\theta_3(q)\theta_4(q)
\end{equation}
\begin{equation}
\frac{\partial}{\partial x}\log \Theta_3(u(y,x)|q(x)) = \frac{yE(x)K(x)+(1-y)(xK(x)^2-2xK(x)\Pi(y,x)-(y-x)\Pi(y,x)^2)}{4yx(1-x)K(x)^2}.
\end{equation}
\begin{equation}
x=\frac{\theta_2(q)^4}{\theta_3(q)^4}, \qquad y = \frac{\theta_2(q)^2}{\theta_3(q)^2} \frac{\Theta_2(u|q)^2}{\Theta_3(u|q)^2}.
\end{equation}
\begin{equation}
1-x=\frac{\theta_4(q)^4}{\theta_3(q)^4},\qquad 1-y = \frac{\theta_4(q)^2}{\theta_3(q)^2} \frac{\Theta_4(u|q)^2}{\Theta_3(u|q)^2}, \qquad y-x = -\frac{\theta_2(q)^2 \theta_4(q)^2}{\theta_3(q)^4} \frac{\Theta_1(u|q)^2}{\Theta_3(u|q)^2}.
\end{equation}

\section{Periods of a degenerate hyperelliptic curve}\label{sec3-geom-int} 
\begin{figure}[h!]\centering
	\includegraphics[width=.9\linewidth]{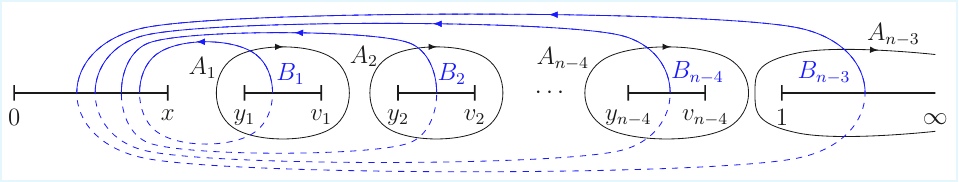}
	\caption{$A$ and $B$ periods on the hyperelliptic curve.}
	\label{fig:PeriodsN}
\end{figure}

Consider a genus $g= n-3$ hyperelliptic curve
\begin{equation}
w^2 = z(1-z)(x-z)\prod_{k=1}^{n-4}(z-y_k)(z-v_k), 
\end{equation}
Choose cycles $A_1, \ldots A_{n-3}$ and $B_1, \ldots, B_{n-3}$ as shown on figure \ref{fig:PeriodsN}. 
\\
The intersection numbers of the cycles are
\begin{equation}\label{eq:cyclesintersections}
A_i\cap B_j = \delta_{ij}, \quad A_i\cap A_j = B_i\cap B_j = 0. 
\end{equation}
On a genus $g$ curve there is a $g$-dimensional space of holomorphic differentials with the following basis:
\begin{equation}
\eta_{n-3} = \frac{dz}{w}\prod_{l=1}^{n-4}(z-y_l), \quad \eta_k = \frac{dz}{w} \prod_{\substack{l=1\\ l\neq k}}^{n-4}(z-y_l), \qquad \text{for}\quad k = 1, \ldots, n-4.
\end{equation}
When $v_k\to y_k$ the curve degenerates and the basis becomes
\begin{equation}
\eta_{n-3} = \frac{dz}{\sqrt{z(1-z)(x-z)}}, \quad \eta_k = \frac{dz}{\sqrt{z(1-z)(x-z)}}\frac{1}{z-y_k} , \qquad \text{for}\quad k = 1, \ldots, n-4.
\end{equation}
The $\eta_{n-3}$ is the unique (up to normalization) holomorphic differential on the degenerate curve and $\eta_k$ have simple poles at the two points of the curve with $z = y_k$. Such differentials are called abelian differentials of the third kind. They can be expressed in terms of theta functions up to terms with holomorphic differentials:
\begin{equation}
\eta_k = \lambda_k \eta_{n-3} +  \frac{dz}{\sqrt{y_k(1-y_k)(x-y_k)}} \frac{\partial}{\partial z} \log \frac{\Theta_1(u(z,x)-u(y_k,x)|q)}{\Theta_1(u(z,x)-\pi + u(y_k,x)|q)}.
\end{equation} 
Here the zeroes of the theta functions correspond to the two poles of the differential.

There is another basis of differentials, which are normalized with respect to the $A$ periods:
\begin{equation}
\int_{A_j}\omega_k = \delta_{jk}.
\end{equation}
Using this basis one defines the period matrix $\tau$ of the curve:
\begin{equation}
\tau_{jk} = \int_{B_j}\omega_k.
\end{equation}
If the basis of cycles satisfies \eqref{eq:cyclesintersections}, the matrix satisfies Riemann bilinear relations, i.\,e. it is symmetric and its imaginary part is positive definite.

For the $\eta$ basis we have
\begin{equation}
	\int_{A_{n-3}}\eta_{n-3} =  4K(x), \quad \int_{A_j}\eta_{n-3} = 0, \qquad \int_{A_j}\eta_k = -2\pi i \frac{1}{\sqrt{y_k(1-y_k)(x-y_k)}}\delta_{jk}
\end{equation}
for $j, k = 1, \ldots, n-4$. Using the periodic properties of $\Theta_1(u|q)$ and the fact that $\Theta_1(-u|q) = -\Theta_1(u|q)$ one can see that:
\begin{equation}
\int_{A_{n-3}}(\eta_k - \lambda_k \eta_{n-3}) =  \frac{2}{\sqrt{y_k(1-y_k)(x-y_k)}} 
\log \frac{\Theta_1( \frac{\pi}2\tau+\frac{\pi}2-u(y_k,x)|q)\Theta_1(\frac{\pi}2\tau-\pi + u(y_k,x)|q)}{\Theta_1(\frac{\pi}2\tau-\frac{\pi}2 + u(y_k,x)|q)\Theta_1(\frac{\pi}2\tau-u(y_k,x)|q)}
= 
0.
\end{equation}
Thus,
\begin{equation}\label{eq:limitdifferentials}
\omega_{n-3} = \frac{1}{4K(x)} \frac{dz}{\sqrt{z(1-z)(x-z)}}, \quad  \omega_k =  -\frac{dz}{2\pi i}\,\frac{\partial}{\partial z} \log \frac{\Theta_1(u(z,x)-u(y_k,x)|q)}{\Theta_1(u(z,x)-\pi + u(y_k,x)|q)}
\end{equation}
for $k=1, \ldots, n-4$. Now we can compute the period matrix. First of all, we see that
the integrals of $\eta_k$ over $B_j$ diverge when $n-3>j\geq k$, because the contour of integration $B_j$ passes through the pole of $\eta_k$. It follows that the diagonal entries of the period matrix are infinite, except for the last one:
\begin{equation}
\tau_{n-3\,n-3} = -\frac{1}{2K(x)}\int_{x}^1 \frac{dz}{\sqrt{z(1-z)(x-z)}} = \frac{iK(1-x)}{K(x)} = \tau(x), \qquad 
\tau_{kk} = \infty,\quad k<n-3.
\end{equation} 
The rest of the last column of the period matrix is:
\begin{equation}
\tau_{j\,n-3} = \int_{B_j} \omega_{n-3} = - \frac{1}{2K(x)}\int_{x}^{y_j} \frac{dz}{\sqrt{z(1-z)(x-z)}}
= 
\frac{2}{\pi}\int_{x}^{y_j}  dz\, \frac{\partial}{\partial z}u(z,x)
=
\frac{2}{\pi} u(y_j,x).
\end{equation}
The remaining entries in the last row are
\begin{equation}
\int_{B_{n-3}}\omega_k =  \frac{1}{2\pi i} \log \frac{\Theta_1(\frac{\pi}{2}\tau-u(y_k,x)|q)^2\Theta_1(u(y_k,x)-\pi|q)^2}{\Theta_1(\frac{\pi}{2}\tau-\pi + u(y_k,x)|q)^2\Theta_1(-u(y_k,x)|q)^2} =  \frac{2}{\pi}u(y_k,x).
\end{equation}
As expected they are the same as in the last column, since the period matrix is symmetric.

Thus, for the degenerate curve we have the following period matrix:
\begin{equation}\label{eq:PeriodMatrix}
\tau(\{y_k\},x) = 
\begin{pmatrix}
	B & \frac{2}{\pi}\mathbf{u}\\
	\frac{2}{\pi} \mathbf{u}^T & \tau(x)
\end{pmatrix}, \qquad 
\mathbf{u} = \begin{pmatrix}
u(y_1, x)\\
\vdots\\
u(y_{n-4}, x)
\end{pmatrix},
\end{equation}
where $B$ is a $n-4\times n-4$ matrix with $\infty$ on the diagonal. Off-diagonal elements of $B$ are
\begin{equation}
B_{jk}= \frac{1}{2\pi i}\log \frac{\Theta_1(u(y_k,x)-u(y_j,x)|q)^2}{\Theta_1(u(y_k,x) + u(y_j,x)|q)^2}.
\end{equation}
Also we see that the diagonal elements of the period matrix diverge in the following way:
\begin{equation} \label{periodmatrixdiv}
B_{kk} = - 2\int_{x}^{y_k-\varepsilon}\omega_k  \sim \frac{1}{\pi i}\log(\varepsilon u'(y_k,x)) + \frac{1}{\pi i} \log \frac{\Theta'_1(0|q)}{\Theta_1(2u(y_k,x)|q)}.
\end{equation}

\bibliographystyle{JHEP}
\newpage
\bibliography{mlg2}
\end{document}